\documentclass[
reprint,
amsmath,
amssymb,
aps,
showkeys
]{revtex4-2}
\usepackage{graphicx}
\usepackage{dcolumn}
\usepackage{bm}
\usepackage{url}
\usepackage{hyperref}
\usepackage{float}
\usepackage{multirow}
\usepackage{mathrsfs}
\usepackage{hyperref}
\hypersetup{colorlinks,allcolors=blue,urlcolor=black}
\usepackage{color,soul}
\usepackage{wasysym}
\usepackage{tikz,lipsum,lmodern}
\usepackage{xurl}
\usepackage{array}

\begin{document}

\title{Google Quantum AI’s Quest for Error-Corrected Quantum Computers}

\author{Muhammad AbuGhanem{$^{1,2}$}}
\address{$^{1}$ Faculty of Science, Ain Shams University, Cairo, 11566, Egypt}
\address{$^{2}$ Zewail City of Science, Technology and Innovation, Giza, 12678, Egypt}.

\email{gaa1nem@gmail.com\\mghanem@sci.asu.edu.eg}

\date{\today}

\begin{abstract}

Quantum computers stand at the forefront of technological innovation, offering exponential computational speed-ups that challenge classical computing capabilities. At the cutting edge of this transformation is Google Quantum AI, a leader in driving forward the development of practical quantum computers. This article provides a comprehensive review of Google Quantum AI's pivotal role in the quantum computing landscape over the past decade, emphasizing their significant strides towards achieving quantum computational supremacy. By exploring their advancements and contributions in quantum hardware, quantum software, error correction, and quantum algorithms, this study highlights the transformative impact of Google Quantum AI's initiatives in shaping the future of quantum computing technology.

\end{abstract}


\maketitle

\tableofcontents

\section{Introduction}

The Noisy Intermediate-Scale Quantum (NISQ) computing era~\citep{NISQ18} has marked a significant milestone in the evolution of quantum technology~\citep{NISQ24}. While these quantum processors are powerful~\citep{art19,Assessing,Light,PhotonicQuantumComputers}, they still face substantial challenges due to escalating error rates, as system sizes increase~\citep{bit-fliperrors}, presenting significant challenges to reliable computation~\citep{DiVincenzo,NISQ18}. 

To address these limitations~\citep{NISQ18}, error-corrected quantum computers are crucial for advancing quantum technology in the post-NISQ era. These systems aim to mitigate the inherent errors associated with qubit operations~\citep{Utility_3,Google_superm_2023}, particularly as system sizes increase, 
enabling more reliable computation. By focusing on reducing operational error rates in quantum processing units (QPUs)~\citep{QREM_0,QREM_1}, we can unlock the full potential of quantum computing~\citep{OnthepowerQC,qc}, paving the way for large-scale quantum computers capable of executing complex, error-corrected computations~\citep{QErrorCorrection1,QErrorCorrection2,GoogleAI2023_2}. This progress is essential for pushing the boundaries of quantum technology and realizing meaningful applications across diverse fields, 
from drug design~\citep{GoogleAI2023_8} to cryptography~\citep{Shor} and beyond~\citep{NISQ24}.

At the forefront of this transformation is Google Quantum AI, a key player pushing the boundaries of what quantum technology can achieve~\citep{art19}. Since its inception, Google Quantum AI has been instrumental in advancing the field, particularly through its innovations in superconducting qubits and its ambitious pursuit of quantum computational supremacy~\citep{qsuperm1,nisqQC10}.

This paper explores Google Quantum AI's pivotal role in advancing quantum computing technology. The research highlights the company's journey from early developments to achieving notable milestones in quantum supremacy. With a focus on their advancements in quantum hardware, quantum software, and quantum error correction (QErC)~\citep{QErrorCorrection1,QErrorCorrection2,GoogleAI2023_2}.

Our exploration offers insights into their impact on both theoretical and practical aspects of quantum computing. 
By analyzing their research and development efforts year by year, we aim to provide a comprehensive overview of how Google Quantum AI has shaped the landscape of quantum technology and what lies ahead in their ongoing quest to build error-corrected, large-scale quantum computers.

The subsequent sections of this paper are structured as follows: 
Section~\ref{sec:overview} presents an overview of Google Quantum AI, highlighting its foundational role and key milestones. 
Section~\ref{sec:yearbyyear} chronicles the company's progress year by year from 2013 to 2024. 
Section~\ref{sec:NISQprocessorsVsclassical} discuss comparing the performance of NISQ processors with classical computers. 
Section~\ref{sec:Supremacy} addresses Google Quantum AI’s achievement of computational supremacy, with detailed examinations of the 53-qubit \textit{Sycamore} processor (SYC-53) and subsequent quantum processors like the SYC 67- and SYC 70-qubit models. 
Section~\ref{sec:Software} provides an overview of Google’s contributions to quantum software and open-source tools, highlighting key platforms such as Cirq, OpenFermion, TensorFlow Quantum, Qsim, and discussing their impact on quantum algorithm development and research. 
Section~\ref{sec:roadmap} discusses Google Quantum AI roadmap and the path towards building large-scale useful quantum computers, emphasizing the strategies and challenges involved. The paper concludes with a summary of findings and reflections on the future trajectory of quantum computing technology in Section~\ref{Sec:Conclusion}.

\section{Google Quantum AI}\label{sec:overview}

\begin{figure*}
    \centering
    \includegraphics[width=0.9\textwidth]{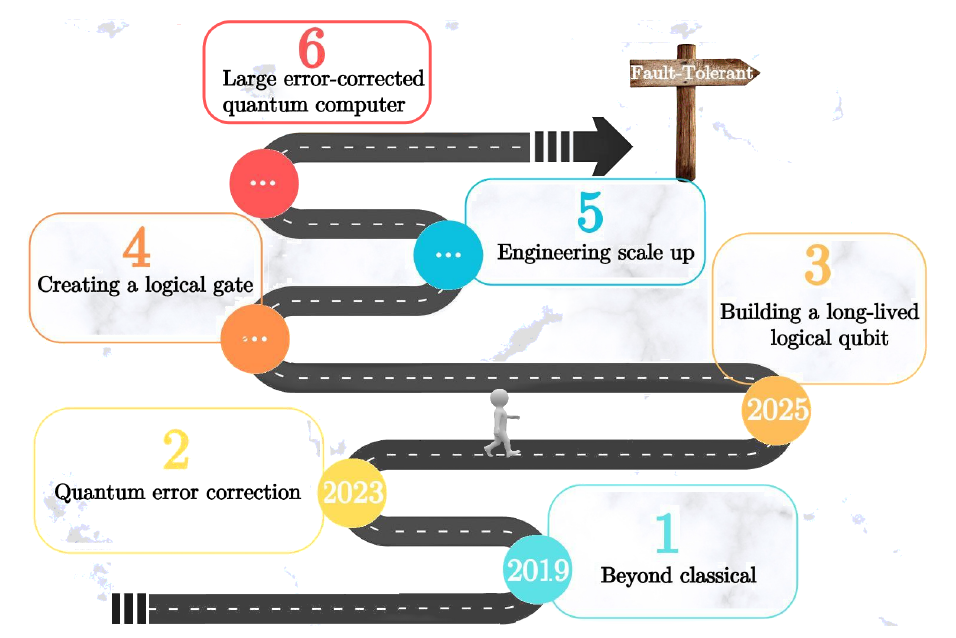}
\caption{The roadmap for quantum computing at Google Quantum AI illustrates their commitment to unlocking the ultimate potential of quantum computing through the development of a large-scale, error-corrected computer. The journey is guided by six pivotal milestones. The first milestone, beyond classical, was achieved in 2019, marking a significant advance over classical computing~\citep{art19}. The second milestone, error-corrected qubits (see Section~\ref{errorcorrection}). Subsequent milestones include building a long-lived logical qubit, creating a logical gate, and engineering scale-up. The final milestone, a large error-corrected quantum computer, represents the ultimate goal of connecting and controlling 1 million qubits, pushing the boundaries of quantum technology to realize meaningful applications.}
    \label{fig:roadmap0}
\end{figure*}

Google Quantum AI stands at the forefront of quantum computing innovation, shaping the future of the field with transformative advancements in technology and theory~\citep{art19}. Since its inception, Google Quantum AI has been dedicated to pushing the boundaries of classical computing, leveraging quantum mechanics to solve complex problems across various domains, including physics~\citep{art19}, chemistry~\citep{GoogleAI8}, and optimization~\citep{VQA_Review,GoogleAI13}.

Central to Google Quantum AI's efforts is the Quantum Artificial Intelligence Lab (QuAIL), a collaborative initiative established in 2013 with partners such as NASA and the Universities Space Research Association (USRA). This lab has been instrumental in advancing quantum computing research, culminating in the landmark achievement of quantum supremacy regime in 2019~\citep{art19}—a milestone that demonstrated the practical capabilities of quantum processors to outperform classical supercomputers on specific tasks~\citep{qsuperm1,NISQ24}.

A cornerstone of Google Quantum AI's research is the development and refinement of superconducting qubits~\citep{wall04,you11,SC_currentstateofplay}, which are implemented using superconducting circuits~\citep{9array2,9array2,Gocircuit1}. These circuits, often based on Josephson junctions, represent a promising approach to scalable quantum computing due to their compatibility with existing semiconductor technologies and their potential for high fidelity and coherence~\cite{dev85,nak97,moo99}.

One of the key achievements of Google Quantum AI in this area has been the significant improvement in qubit coherence times~\citep{GoogleAI2023_1}. Coherence time, the period during which a qubit maintains its quantum state without succumbing to decoherence, is crucial for reliable quantum computations. By addressing environmental noise and enhancing qubit design, Google Quantum AI has successfully extended coherence times, thereby enhancing the stability and reliability of their qubits.

Error management remains another critical focus for Google Quantum AI. Quantum systems are inherently susceptible to errors from noise, control imperfections, and environmental interactions~\citep{QErrorCorrection1,QErrorCorrection2}. To address these challenges, the team has made notable progress in reducing error rates through sophisticated error correction techniques and optimization of qubit designs~\citep{GoogleAI2023_1,Gocircuit2,Gograph2}. These advancements contribute to more accurate and stable quantum computations.

Google Quantum AI has utilized their quantum processors to explore novel physical phenomena. This includes observing time crystals~\citep{time-crystalline} and Majorana edge modes~\citep{Majorana}, as well as making new experimental discoveries like robust bound states of interacting photons~\citep{boundstates} and the noise resilience of Majorana edge modes in Floquet evolutions~\citep{Majorana}.

Looking ahead, Google Quantum AI is committed to further advancing qubit performance and expanding its quantum infrastructure. Their ambitious goal is to construct a fully fault-tolerant quantum computer by integrating thousands of surface-encoded logical qubits, significantly reducing error rates and overcoming the limitations of current technology~\citep{surfacecode}. 

As the field progresses, Google Quantum AI's ongoing contributions will undoubtedly continue to shape the future of quantum computing and drive the next wave of technological innovation.

\section{A Decade of Research and Innovation}\label{sec:yearbyyear}

In this section, we shed light on the progress and achievements of Google Quantum AI over the past decade, tracing their journey from early developmental stages to achieving quantum computational supremacy and advancing QErC (see Figure~\ref{fig:roadmap0}). As a leader in the field, Google Quantum AI has made significant strides, and we will delve into key research from each year to highlight their major contributions and their impact on the broader quantum computing landscape.

\subsection{2013: Early developments}

The application of quantum computing to machine learning tasks was explored in 2013~\citep{GoogleAI1}. The study examined how quantum algorithms could potentially enhance the efficiency of processing and classifying large datasets within high-dimensional spaces~\citep{Mackay_2}. Unlike classical algorithms, which typically exhibit polynomial time complexity, these quantum algorithms are capable of performing clustering tasks in logarithmic time relative to the number of data points and dimensions, thereby offering an exponential speed-up.

\subsection{2014: Defining and detecting quantum speedup}

In 2014, a superconducting qubit architecture featuring high-coherence qubits and dynamically tunable coupling is presented in~\citep{GoogleAI5}. This architecture (known as \textit{gmon}, constructed using the \textit{Xmon} transmon design~\citep{Xmondesign_5}) allows the coupling to be set to zero and adjusted with nanosecond precision, thereby avoiding frequency crowding issues and supporting a range of applications, including quantum logic gates and simulations. The approach was demonstrated with a novel, fast adiabatic controlled-Z gate (CZ), showcasing potential for scalable quantum computation.

The research in~\citep{GoogleAI6} focused on defining and measuring quantum speedup. Using data from a D-Wave Two device with up to 503 qubits, the study conducted benchmark tests with random spin glass instances~\citep{Spinglass_37}. The results showed no clear evidence of quantum speedup, both when analyzing the full dataset and its subsets.

The first experimental evidence of the computational role of multi-qubit quantum tunneling in quantum annealers~\citep{quantum annealers_4,quantum annealers_5} is presented in~\citep{GoogleAI2}. Utilizing the NIBA (Non-interacting Blip Approximation) Quantum Master Equation~\citep{NIBA_26}, the researchers developed a model that describes the effects of multi-qubit dissipative tunneling in the presence of complex noise. The model's predictions align well with experimental data from the D-Wave Two quantum annealer, illustrating an increase in success probabilities for problems with up to 200 qubits.

Quantum coherence in scalable quantum annealing processors using qubit tunneling spectroscopy~\citep{tunnelingspectroscopy_33} is explored in~\citep{GoogleAI5_1}. The study revealed that qubits in both two- and eight-qubit systems became entangled during quantum annealing, and this entanglement persisted even at thermal equilibrium.

Advanced superconducting quantum circuits are leveraged in~\citep{GoogleAI3}, to explore the topological properties of quantum systems. By measuring the deflection of quantum trajectories in the parameter space of a Hamiltonian~\citep{Hamiltonian_6}, the researchers applied a quantum analog of the Gauss–Bonnet theorem, revealing essential topological features. They validated their technique using the Haldane model~\citep{Haldanemodel_7} and extended it to interacting systems with a certain qubit architecture~\citep{newqubit_8,newqubit_9}, uncovering an interaction-induced topological phase. This approach established a robust platform for studying topological phenomena in quantum systems.

Dynamics-based sampling methods are improved~\citep{GoogleAI4}, by introducing variables that stabilized momentum fluctuations caused by noise from stochastic gradients~\citep{stochastic_9,stochastic_5}. Drawing inspiration from statistical mechanics~\citep{StatisticalMechanics_23}, this method enhanced sampling efficiency and stability in large datasets.

\subsection{2015: Quantum hardware design and error correction}

In 2015, a tunable coupler design for superconducting Xmon qubits was investigated in~\citep{GoogleAI7}, utilizing a flux-biased Josephson junction as a tunable current divider. The study calculated the effective qubit-qubit interaction Hamiltonian, finding that the qubit's nonlinearity reduces the transverse coupling by about $15\%$ and introduces a small diagonal coupling. The approach offers insights applicable to other complex nonlinear circuits in quantum hardware design~\citep{Xmondesign_5}.

The simulation of the bond dissociation curve of the helium hydride cation (HeH$^+$ is thought to be the first molecule formed in the early universe~\citep{HeH+}) using a solid-state quantum register based on nitrogen-vacancy (NV) centers in diamond~\citep{NVcenters_44} was demonstrated in~\citep{GoogleAI8}. The researchers achieved an energy uncertainty of about $10^{-14}$ Hartree, surpassing the precision required for chemical applications. This work marks significant progress toward scalable quantum chemistry simulations using solid-state quantum platforms~\citep{qsimulations_Feynman,qsimulations_Lloyd,qsimulations_2019}.

Advancements in QErC were highlighted in~\citep{GoogleAI9}, where a nine-qubit linear array was used to protect classical states from bit-flip errors~\citep{bit-fliperrors}. The researchers demonstrated reduced error rates with increasing system size, achieving a factor of 2.7 improvement with five qubits and 8.5 with nine qubits after eight cycles. Additionally, the preservation of a non-classical Greenberger–Horne–Zeilinger (GHZ) state~\citep{GHZ} was verified, 
and effectively reducing environment-induced errors. Additional research efforts from the same year can be found in~\citep{GoAI2015_1,GoAI2015_2,GoAI2015_3}.

\subsection{2016: Quantum algorithms and system performance}

In 2016, the variational quantum eigensolver (VQE) algorithm~\citep{similarity_transf_31} was improved in~\citep{GoogleAI2016_1} by introducing a variational adiabatic ansatz, quantum variational error suppression, and cost-reducing techniques such as truncation. The study demonstrated that modern optimization methods can significantly lower computational costs.

Digitized adiabatic quantum computing~\citep{Adiabatic_14,Adiabatic_15} was implemented in a superconducting system~\citep{GoogleAI12}. This study combined adiabatic~\citep{adiabaticquanumcomputing_3,adiabaticquanumcomputing_4,adiabaticquanumcomputing_5} and digital quantum computing~\citep{qsimulations_Lloyd,digitalqsimulation_9,digitalqsimulation_10,digitalqsimulation_11,digitalqsimulation_12}, using tomography to probe the system’s evolution and assess error scaling with system size~\citep{bit-fliperrors}. The researchers solved the one-dimensional Ising problem and more complex Hamiltonians with up to nine qubits and 1,000 gates, contributing to the development of scalable quantum computing systems.

The first electronic structure calculation on a quantum computer without extensive pre-compilation was performed in~\citep{GoogleAI13}. Using superconducting qubits, the researchers computed the energy surface of molecular hydrogen using two quantum algorithms: the UCC (unitary coupled cluster)~\citep{unitarycoupledclustertheory_20,unitarycoupledclustertheory_21,unitarycoupledclustertheory_22,coupledcluster_39} method via the VQE~\citep{similarity_transf_31} and Trotterization~\citep{Trotterization_34} with quantum phase estimation (QPE)~\citep{QPE_35}. The VQE achieved chemical accuracy and demonstrated better robustness to errors compared to traditional methods, highlighting its potential for future simulations of complex molecules.

Ergodic dynamics in a 3-qubit superconducting system were demonstrated in~\citep{GoogleAI11}. The research showed that the system explores all states over time, similar to classical chaotic systems~\citep{chaotic systems_18,chaotic systems_14,chaotic systems_15,chaotic systems_17,chaotic systems_16}. By measuring entanglement entropy, the study revealed that the system acts as a reservoir, increasing entropy through entanglement. For additional research efforts by Google Quantum AI in 2016, readers are directed to~\citep{GoogleAI10,GoAI2016_1,GoAI2016_2,GoAI2016_3,GoAI2016_4,GoAI2016_5,GoAI2016_6}.

\subsection{2017: Commercialize quantum technologies in five years}

In 2017, Google's Quantum AI outlined strategies for commercializing quantum technologies within five years~\citep{GoogleAI18}. Focusing on investment opportunities and steps needed to advance towards fully functional quantum machines, providing a roadmap to accelerate the development and deployment of these technologies (see Section~\ref{sec:roadmap}).

A scalable trapped ion quantum computer~\citep{Ion-trap_8,Ion-trap_9,Ion-trap_10,Ion-trap_11} was proposed in~\citep{GoogleAI19}, featuring a modular design based on silicon microfabrication and long-wavelength radiation-based quantum gates~\citep{gatemechanism_12,gatemechanism_13}. The design supports fault-tolerant operations through high error-threshold surface codes and allows for interconnection of standalone units via ion transport~\citep{iontransport_18,iontransport_16,iontransport_17}. This approach can be adapted for various quantum computing architectures, including those utilizing photonic interconnects~\citep{PhotonicQuantumComputers,Light}.

Chiral ground-state currents of interacting photons were demonstrated in~\citep{GoogleAI20} using superconducting qubits in synthetic magnetic fields. By modulating qubit couplings to create artificial magnetic fields, the researchers observed directional photon circulation and strong photon interactions~\citep{photonhopping_33,photonhopping_24,photonhopping_34}. This setup offers a new platform for exploring quantum phases in systems with strongly interacting photons.

The energy levels of interacting photons were analyzed using a chain of nine superconducting qubits in~\citep{GoogleAI21}. This study revealed features of the Hofstadter butterfly spectrum~\citep{Hofstadterbutterfly_25,Hofstadterbutterfly_28} and observed a transition from a thermalized to a localized phase by introducing disorder. This technique provides a novel approach to studying quantum phases of matter through many-body spectroscopy~\citep{Many-bodylocaliz_10,Many-bodylocaliz_8,Many-bodylocaliz_9}. Additional research efforts from the same year can be found in~\citep{GoogleAI14,GoogleAI15,GoogleAI16,GoogleAI17,GoAI2017_1,GoAI2017_2,GoAI2017_3,GoAI2017_4,GoAI2017_5}.

\subsection{2018: The path to quantum supremacy with superconducting qubits}

In 2018, several key contributions advanced the quest for quantum supremacy and improved quantum simulations~\citep{qsimulations_Feynman,qsimulations_Lloyd,qsimulations_2019}. The investigation of unpredictable fluctuations in energy-relaxation times of superconducting qubits was performed in~\citep{GoogleAI24}. This study identified individual two-level-system defects as the primary cause of these fluctuations by using qubits as temporal and spectral probes. The findings provide a basis for enhancing qubit stability through improved calibration, design, and fabrication techniques.

A dual plane wave basis for quantum simulation of electronic structures was introduced in~\citep{GoogleAI25}. This method reduces the complexity of Hamiltonians from ${\cal O}(N^4$) to ${\cal O}(N^2$) terms, facilitating more efficient Hamiltonian steps and reducing circuit depths for Trotter~\citep{Trotter,Trotter1} and Taylor-series simulations~\citep{Taylor-simu}. It also requires fewer measurements for variational algorithms.

The simulation of the electronic structure Hamiltonian using a "fermionic swap network" was presented in~\citep{GoogleAI26}. This approach achieves Trotter steps with linear depth ($N$) and requires ${\cal O}(N^2$) two-qubit entangling gates~\citep{Entangling,SQSCZ1,SQSCZ2}. It also allows for the preparation of Slater determinants with a depth of at most $N/2$, even on minimally connected qubit architectures. The authors suggest that this method is optimal in terms of gate count for Trotter steps and represents a significant practical advancement for quantum chemistry simulations~\citep{GoogleAI2018_2}.

A strategy for achieving quantum computational supremacy using superconducting qubits was outlined in~\citep{GoogleAI27}. By tuning nine qubits to generate diverse Hamiltonian evolutions, the researchers observed output probabilities aligning with a universal distribution, indicating uniform sampling of the Hilbert space~\citep{GoogleAI30}. These results demonstrate that increasing the qubit count could allow quantum systems to solve problems beyond the reach of classical computers.

The application of the VQE algorithm~\citep{similarity_transf_31} to molecular energy simulations using the UCC (unitary coupled cluster) ansatz~\citep{similarity_transf_31,GoogleAI2016_1,unitarycoupledclustertheory_20} was explored in~\citep{GoogleAI28}. The study introduced methods to reduce circuit depth and enhance wave-function optimization through efficient classical approximations of cluster amplitudes~\citep{classicalcoupledcluster_41,classicalcoupledcluster_42,classicalcoupledcluster_43,classicalcoupledcluster_44}.

The challenges of using random circuits as initial guesses for hybrid quantum-classical algorithms~\citep{GoogleAI2016_1} were revealed in~\citep{GoogleAI2018_1}. They showed that the probability of obtaining a non-zero gradient decreases exponentially with the number of qubits, highlighting the need for improved designs of parameterized quantum circuits.

A QNN (quantum neural network) for binary classification tasks was presented in~\citep{GoogleAI29}. This QNN handles both classical and quantum data by applying parameterized unitary transformations to an input quantum state and measuring a Pauli operator to predict binary labels. The study also explored adapting the QNN for quantum data, suggesting its suitability for near-term quantum processors~\citep{NISQ18}.

Sampling from RQCs as a task to demonstrate quantum supremacy~\citep{qsuperm1} was proposed in~\citep{GoogleAI30}. The study argued that classical computation would require exponential time for this task, and introduced XEB~\citep{GoogleAI30,GoogleAI27,art19,XEB18} to measure experimental fidelity in complex multi-qubit dynamics. It found that a $7\times7$ qubit grid performing around 40 clock cycles with specific error rates could achieve this milestone~\citep{NISQ24}.

The experimental implementation of the VQE algorithm~\citep{similarity_transf_31} using a trapped-ion quantum simulator~\citep{Trapped-ionquantumsimulator} to calculate the ground-state energies of two molecules was reported in~\citep{GoogleAI2018_2}. The study demonstrated and compared different encoding methods with up to four qubits, addressed measurement noise and its mitigation, and explored adaptive strategies for achieving chemical accuracy, setting a benchmark for multi-qubit quantum simulators~\citep{QuantumSimulators_1,QuantumSimulators_2,QuantumSimulators_3,QuantumSimulators_4}. Additional research efforts from the same year can be found in~\citep{GoogleAI22,GoogleAI23,GoAI2018_1,GoAI2018_2,GoAI2018_3,GoAI2018_4,GoAI2018_5,GoAI2018_6,GoAI2018_7,GoAI2018_8,GoAI2018_9,GoAI2018_10,GoAI2018_11}.

\subsection{2019: Quantum supremacy—the \textit{Sycamore} processor and beyond}

\begin{figure*}
    \centering
    \includegraphics[width=\linewidth]{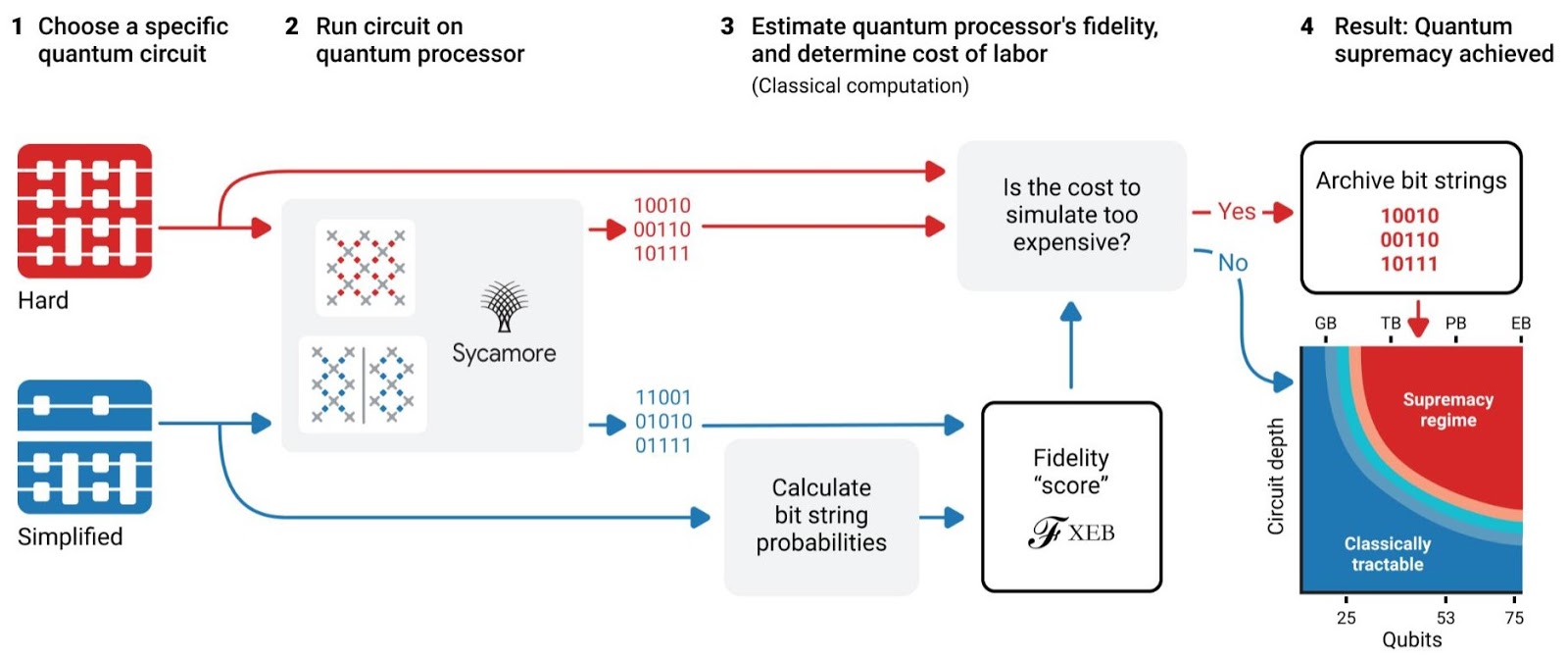}
\caption{Demonstration of quantum computational supremacy using SYC-53. The process comprises four steps: First, a specific quantum circuit is selected. Next, the circuit is executed on the quantum processor. Following this, the processor's fidelity is estimated, and the labor cost is assessed. Finally, the results indicate that quantum computational supremacy has been achieved. Reproduced from~\citep{Supremacy19}.}
    \label{fig:supremacySYC-53}
\end{figure*}

In 2019, significant advancements in quantum computing were achieved. Quantum supremacy regime~\citep{qsuperm1,nisqQC10} was demonstrated in~\citep{art19} using a superconducting qubit processor with 54 qubits (see Figure~\ref{fig:supremacySYC-53}), creating states in a computational space of dimension \(2^{53}\) (approximately \(10^{16}\)).  Google Quantum AI and Collaborators revealed that the SYC-53 processor sampled quantum states in about 200 seconds, a task that would take approximately 10,000 years on a state-of-the-art classical supercomputer (see Section~\ref{RCS}). Shortly after the publication, a debate arose concerning the possible overestimation of the time needed to solve the same problem on a supercomputer~\citep{Cupjin,feng22,Pednault}. Nevertheless, this achievement stands as a significant milestone in the field~\citep{NISQ24}.

A quantum algorithm for simulating quantum chemistry was introduced in~\citep{GoogleAI33}, with gate complexity scaling as 
\({\cal O}(N^{1/3} \eta^{8/3})\), where \(N\) is the number of orbitals and \(\eta\) is the number of electrons. This method employs interaction picture techniques in the rotating frame of the kinetic operator, offering significant efficiency improvements over previous algorithms~\citep{GoogleAI33_40,GoogleAI33_10,GoogleAI33_9}, particularly for large basis sizes or without the Born-Oppenheimer approximation~\citep{Born-Oppenheimer}.

Diabatic two-qubit gates for frequency-tunable superconducting qubits were reported in~\citep{GoogleAI34}, achieving Pauli error rates as low as \(4.3 \times 10^{-3} \pm 0.0002\) in 18 ns, with average gate fidelities up to \(0.9966 \pm 0.0002\). This improvement is attributed to the synchronization of gate parameters with minima in the leakage channel, which enhances gate robustness. This method, validated with CPHASE~\citep{CPHASE_16} and iSWAP-like~\citep{iSWAP-like} gates, lays the groundwork for extending to multi-body operations.

A prototype cryogenic CMOS quantum controller was presented in~\citep{GoogleAI35}, designed in a 28-nm CMOS process to manage transmon qubits~\citep{koch07,Stef14} with a 16-word XY gate instruction set. The study details the controller's design and performance, noting low error rates, scalability, and cryogenic operation. It achieves under 2 mW power dissipation and a digital data stream of less than 500 Mb/s.

The flexible quantum circuit simulator, qFlex, was introduced in~\citep{GoogleAI36}. Using tensor networks, qFlex computes both exact and low-fidelity amplitudes for verifying and mimicking NISQ devices~\citep{NISQ24,NISQ18}. It efficiently simulates RQCs expected in supremacy experiments~\citep{GoogleAI30,RCS-5,RCS-31,RCS-BFNV18,RCS-Mov18,RCS2019,RCS-11,GoSup23_21,GoSup23_19,RCS2023,RCSComp2023}, achieving simulations at a cost $1/f$ of perfect fidelity ones. The study also eliminates rejection sampling overhead and benchmarks qFlex on NASA HPC clusters~\citep{NASAHPC}, achieving a peak performance of 20 PFLOPS, the highest numerical computation in terms of sustained floating-point operations per second (FLOPS) and node usage on these clusters.

The improvement of QNNs through classical neural networks and meta-learning, alternatively referred to as ``learning to learn"~\citep{learningtolearn_38,learningtolearn_35,learningtolearn_34,learningtolearn_36,learningtolearn_37}, was explored in~\citep{GoogleAI37}. By training classical recurrent neural networks to find approximate optimal parameters for quantum algorithms. They investigated the effectiveness of this method for several problem classes, including QAOA for Sherrington-Kirkpatrick Ising models~\citep{GoAI2017_1}, QAOA for MaxCut~\citep{QAOA}, and a VQE for the Hubbard model~\citep{unitarycoupledclustertheory_22,GoogleAI26,GoAI2018_3}. The approach reduces the number of optimization iterations needed, improves convergence to local minima, and generalizes well across different problem sizes.

A framework combining reinforcement learning (RL) and deep neural networks (DNNs)~\citep{DRL_15,DRL_12,DRL_13,DRL_14,DRL_16} to optimize quantum control was introduced in~\citep{GoogleAI38}. The framework (namely ``Universal control cost Function Optimization" or ``UFO") focuses on improving both speed and fidelity of two-qubit gates by incorporating control noise into the training environment~\citep{fullcontrollability_20,fullcontrollability_21,fullcontrollability_22}. This approach reduces average gate error~\citep{minimizeresonantleakage_1,minimizeresonantleakage_28,minimizeresonantleakage_29,minimizeresonantleakage_30,minimizeresonantleakage_31} by two orders of magnitude and gate time by up to one order of magnitude compared to traditional methods~\citep{Dauphin_35}, offering promising applications for quantum simulation~\citep{QuantumSimulators_3}, quantum chemistry~\citep{GoogleAI2018_2}, and quantum supremacy~\citep{NISQ18,NISQ24} tests with near-term quantum devices~\citep{GoogleAI30,NISQ24}. Additional research efforts from the same year can be found in~\citep{GoogleAI31,GoogleAI32,GoAI2019_1,GoAI2019_2,GoAI2019_3,GoAI2019_4,GoAI2019_5,GoAI2019_6}.

\subsection{2020: Quantum chemistry and simulation techniques}

In 2020, several notable advancements were witnessed. Quantum simulations of chemistry, including diazene isomerization, were explored in~\citep{OpenFermion} using up to twelve qubits. This paper introduces error-mitigation techniques based on “$N$-representability"~\citep{n-representability_31,n-representability_32,n-representability_33} to enhance fidelity. By employing parameterized ansatz circuits and variationally optimizing Givens rotation for non-interacting fermion evolution, the study prepares the Hartree-Fock wave function~\citep{Hartree-Focktheory}. Although this method remains classically tractable, it generates highly entangled states, offering insights into hardware performance and paving the way for larger-scale quantum chemistry simulations
~\citep{qsimulations_Feynman,qsimulations_Lloyd,qsimulations_2019,qchemistry_12,qchemistry_16,qchemistry_7,qchemistry_39,qchemistry_32}.

The introduction of OpenFermion, an open-source Python library for quantum simulations, is detailed in~\citep{OpenFermion0}. This library simplifies the simulation of fermionic and bosonic systems, providing an interface to electronic structure packages. It facilitates the translation from molecular specifications to quantum circuits, reducing the need for deep domain expertise (see section~\ref{sec:Software}).

TensorFlow Quantum (TFQ), an open-source library for designing and training hybrid quantum-classical models, was introduced in~\citep{TensorFlow_whitepaper}. TFQ integrates with TensorFlow to support quantum circuit simulations and offers high-level tools for tasks such as quantum classification, control, and optimization. It also facilitates advanced quantum learning applications, including meta-learning~\citep{learningtolearn_38,learningtolearn_35,learningtolearn_34,learningtolearn_36,learningtolearn_37} and VQEs, aiming to enable research in quantum computing and machine learning~\citep{Quantummachinelearning} (see section~\ref{sec:Software}).

The variational quantum unsampling protocol was introduced in~\citep{GoogleAI40}. This method learns the features of quantum circuits by unraveling their unknown dynamics using a nonlinear QNN. Tested on a photonic quantum  processor~\citep{PhotonicQuantumComputers}, the protocol aids in verifying quantum outputs and has broad applications, including quantum measurement~\citep{Quantummeasurement}, quantum tomography~\citep{QPT,SQSCZ2}, sensing~\citep{Quantumsensing}, quantum imaging~\citep{quantumimaging}, and ansatz validation~\citep{unitarycoupledclustertheory_22,ansatzvalidation_39,ansatzvalidation_40}.

A 16-qubit digital superconducting quantum processor was used in~\citep{GoogleAI43} to simulate the dynamics of the one-dimensional Fermi-Hubbard model~\citep{Fermi-Hubbardmodel_1,Fermi-Hubbardmodel_2,Fermi-Hubbardmodel_16,Fermi-Hubbardmodel_26,QC_Fermi-Hubbard}, revealing distinct spreading velocities of charge and spin densities in the highly excited regime. Advanced gate calibration and error-mitigation techniques~\citep{GoogleAI43_12,GoogleAI43,GoogleAI43_32,GoogleAI43_33} were introduced to address systematic errors and decoherence, enabling accurate simulation despite complex circuits.

The paper~\citep{GoogleAI47} presents qFlex, a tensor-network-based classical simulator for benchmarking NISQ  computers~\citep{NISQ24}. The study details high-performance computing simulations of RQCs on \textit{Summit}, the world’s fastest supercomputer (at that time), achieving 281 PFLOP/s. The results highlight the significant energy efficiency advantage of NISQ devices~\citep{NISQ18,NISQ24} over classical supercomputers and propose a standard benchmark for evaluating NISQ systems based on qFlex.

A continuous two-qubit gate set for gmon qubits was demonstrated in~\citep{GoogleAI48}, achieving a threefold reduction in circuit depth compared to standard methods. The implemented gates include an iSWAP-like gate~\citep{iSWAP-like} for arbitrary swap angles and a controlled-phase gate for arbitrary conditional phases~\citep{GoogleAI34}, enabling the full fSim gate set. Results show an average two-qubit Pauli error of \(3.8 \times 10^{-3}\), reflecting high fidelity across 525 different fSim gates. Additional research efforts from the same year can be found in~\citep{GoogleAI41,GoogleAI42,GoogleAI44,GoogleAI45,GoogleAI46,GoogleAI39,GoAI2020_1,GoAI2020_2,GoAI2020_3,GoAI2020_4,GoAI2020_5}.

\subsection{2021: Quantum machine learning and optimization}

In 2021, Google Quantum AI made significant progress in various areas of quantum computing. 
The application of quantum computing to machine learning was explored in~\citep{GoogleAI2021_1}. 
The method to factor 2048-bit RSA integers in 8 hours using 20 million noisy qubits is presented in~\citep{GoogleAI2021_2}. The paper outlines the requirements for this construction, including logical qubits, Toffoli gates~\citep{Toffoli,Toffoli_ECR}, and measurement depth, and discusses its implications for RSA encryption~\citep{RSA_73} and discrete logarithm-based cryptographic schemes~\citep{QAlg_Shor}.

The use of Google Sycamore’s quantum processor, comprises a 2-dimensional grid of 54 transmon qubits~\citep{art19}, for combinatorial optimization with the QAOA is demonstrated in~\citep{GoogleAI2021_3}. The study explores both hardware-native problems (defined on the connectivity graph of the hardware) and non-native ones, such as the Sherrington-Kirkpatrick model and MaxCut~\citep{Sherrington-Kirkpatrick}. For native problems, performance improves with circuit depth, while for non-native problems, it decreases with size. Although QAOA~\citep{QAOA} circuits outperform random guessing, they do not surpass some classical algorithms, highlighting the difficulty of scaling QAOA~\citep{QAOA} for non-native problems and suggesting a focus on non-native problems for benchmarking quantum processors.

The preparation of the ground state of the toric code Hamiltonian using a superconducting quantum processor is reported in~\citep{GoogleAI2021_5}. The paper measures topological entanglement entropy close to the expected value and performs anyon interferometry to analyze the braiding statistics of emergent excitations~\citep{GoogleAI2021_5_35,GoogleAI2021_5_24,GoogleAI2021_5_25,GoogleAI2021_5_41,GoogleAI2021_5_42}. These results highlight the potential of quantum processors to advance understanding in topological quantum matter and QErC~\citep{GoogleAI2021_5_28,GoogleAI2021_5_29,GoogleAI2021_5_30}.

The use of one-dimensional repetition codes on a 2-dimensional grid of superconducting qubits to achieve exponential suppression of bit-flip and phase-flip errors is demonstrated in~\citep{GoogleAI2021_6}. The study shows that increasing the number of qubits from 5 to 21 reduces logical errors by over 100 times, with stability maintained over 50 rounds of error correction~\citep{QErrorCorrection1,QErrorCorrection2,QErrorCorrection_3}. The paper presents a method for precise error correlation analysis and performs error detection with a small logical qubit using the 2D surface code~\citep{2Dsurfacecode_18,2Dsurfacecode_19}.

An overview of variational quantum algorithms (VQAs) as a promising approach to leveraging quantum computing despite current device limitations is provided in~\citep{VQA_Review}. The rapid development and potential of quantum simulators are discussed in~\citep{GoogleAI2021_4}. The paper highlights the use of entanglement and many-particle behavior to tackle complex scientific and engineering problems, noting that over 300 quantum simulators worldwide have emerged in the last two decades~\citep{GoogleAI2021_4}. Recent advancements promise a new era for both specialized and programmable simulators, recommending investment to enhance scientific applications and foster multi-disciplinary collaborations. Additional research efforts from the same year can be found in~\citep{GoAI2021_1,GoAI2021_2,GoAI2021_3,GoAI2021_4,GoAI2021_5,GoAI2021_6,GoAI2021_7,GoAI2021_8,GoAI2021_9,GoAI2021_10,GoAI2021_11,GoAI2021_12,GoAI2021_13,GoAI2021_14,GoAI2021_15,GoAI2021_16,GoogleAI51,GoAI2021_17,GoogleAI50,GoAI2021_18,GoogleAI49,GoAI2021_19,GoAI2021_20,GoAI2021_21,GoAI2021_22,GoAI2021_23,GoAI2021_24,GoAI2021_25,GoAI2021_26,GoAI2021_27,GoAI2021_28,GoAI2021_29,GoAI2021_30}.

\subsection{2022: Quantum advantage in learning from experiments}

In 2022, the potential of quantum technology to significantly enhance learning from experiments is demonstrated in~\citep{GoogleAI2022_1}. The study reveals that quantum machines can offer exponential advantages in predicting system properties, performing quantum principal component analysis on noisy states, and learning physical dynamics~\citep{Quantumsensing,Quantummachinelearning}. Experimental results using up to 40 superconducting qubits and $1,300$ quantum gates confirm that substantial advantages are achievable even with current noisy quantum processors~\citep{NISQ18}.

The observation of a discrete time crystal (DTC)~\citep{DTC_13,DTC_14,DTC_15,DTC_9,DTC_7,DTC_10,DTC_11,DTC_12} in a periodically driven many-body-localized (MBL)~\citep{MBL_16,MBL_17} system using superconducting qubits and CPHASE gates is reported in~\citep{GoogleAI2022_2}. The study demonstrates the DTC’s characteristic spatiotemporal response from various initial states and employs a time-reversal protocol to measure external decoherence effects. Efficient sampling of the eigen-spectrum and a finite-size analysis identify the phase transition out of the DTC, showcasing a scalable approach for exploring non-equilibrium phases of matter with quantum processors.

The investigation into the QAOA for the Sherrington-Kirkpatrick model~\citep{Sherrington-Kirkpatrick}, a complex optimization problem with random signed couplings among spins, is detailed in~\citep{GoogleAI2022_3}. The research assesses whether QAOA can match the performance of a classical algorithm by Montanari~\citep{Montanari}, which can efficiently approximate the Sherrington-Kirkpatrick model’s ground state energy. Additional contributions from 2022 include~\citep{GoAI2022_1,GoAI2022_2,GoAI2022_3,GoAI2022_4,GoAI2022_5,GoAI2022_6,GoAI2022_7,GoAI2022_8,GoAI2022_9,GoAI2022_10,GoAI2022_11,GoAI2022_12,GoAI2022_13,GoAI2022_14,GoAI2022_15,GoAI2022_nonu,GoAI2022_16,GoAI2022_17,GoAI2022_18,GoAI2022_19,GoAI2022_20,GoAI2022_21,GoAI2022_22,GoAI2022_23,GoAI2022_24,GoAI2022_25,GoAI2022_26,GoAI2022_27,GoAI2022_28}.

\subsection{2023: Quantum error correction}\label{errorcorrection}

\begin{figure*}
    \centering
    \includegraphics[width=0.75\textwidth]{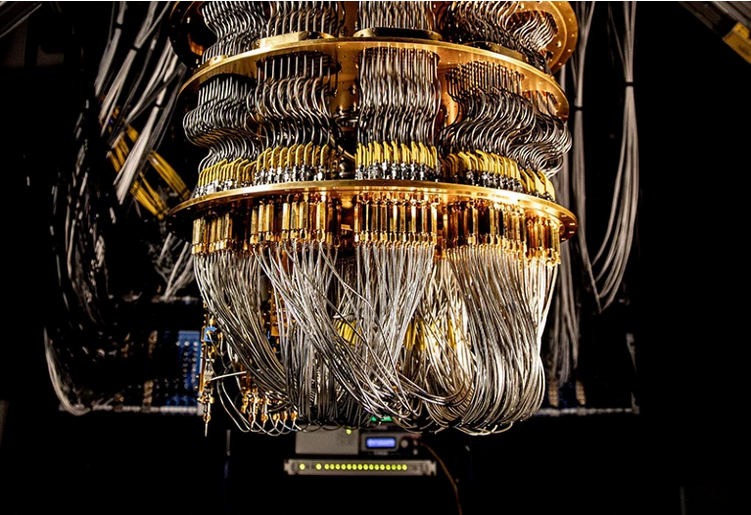}
    \caption{Google’s quantum computer achieves a significant milestone by decreasing error rates. Researchers have shown for the first time that increasing the number of qubits can reduce the error rate in quantum calculations~\citep{GoogleAI2023_1}. Image source~\citep{Google_image}. Credit: Google Quantum AI.}
    \label{fig:Google_image}
\end{figure*}

In 2023, several notable advancements were made, particularly in quantum error correction and achieving quantum computational advantage. The study~\citep{Google_superm_2023} addressed the challenge of incoherent noise in quantum processors, which disrupts long-range correlations and coherent computation. By employing RCS~\citep{GoogleAI30,RCS-5,RCS-31,RCS-BFNV18,RCS-Mov18,RCS2019,RCS-11,GoSup23_21,GoSup23_19,RCS2023,RCSComp2023} and XEB~\citep{GoogleAI30,GoogleAI27,art19,XEB18}, the researchers identify key phase transitions related to cycle counts and error rates per cycle. Their RCS experiment, involving 67 qubits and 32 cycles, demonstrates that quantum processors can reach a computationally complex phase despite noise, surpassing current classical supercomputers' capabilities and marking a significant advancement in handling quantum noise (as detailed in Section~\ref{subsec:SYC67SYC70}).

The performance of logical qubits in a superconducting qubit system with QErC~\citep{QErrorCorrection1,QErrorCorrection2} is reported in~\citep{GoogleAI2023_1}. QErC is capable of significantly reducing operational error rates in quantum processing units, albeit with increased demands on both time and qubit resources~\citep{QuErrCorr_18,QuErrCorr_19}. Numerous studies have successfully implemented error correction using codes designed to rectify single errors, including color QErC code~\citep{QuErrCorr_21}, the distance-3 Bacon–Shor QErC code~\citep{QuErrCorr_20}, 5-qubit QErC code~\citep{QuErrCorr_22}, surface QErC codes~\citep{QuErrCorr_24,QuErrCorr_25}, and heavy-hexagon QErC code~\citep{QuErrCorr_23}, along with (CV) continuous variable QErC codes~\citep{QuErrCorr_26,QuErrCorr_27,QuErrCorr_28,QuErrCorr_29}. Nevertheless, a pivotal question persists: will scaling up the size of the error-correcting code lead to lower logical error rates in practical devices~\citep{GoogleAI2023_1}? 
An answer to this question is reported by Google Quantum AI in~\citep{GoogleAI2023_1}. The study shows that increasing the number of physical qubits enhances logical qubit performance, even though it introduces additional error sources. They unveiled a 72-qubit superconducting processor that employs a 49-qubit distance-5  surface code slightly surpasses the performance of its average subset, a 17-qubit distance-3 surface code (see Figure~\ref{fig:Google_image}). The research also identifies high-energy events affecting performance using a distance-25 repetition code, highlighting progress and ongoing challenges in achieving low logical error rates for practical quantum computing~\citep{High-fidelity_trapped-ion,High-fidelity_silicon,High-fidelity_near-term,bit-fliperrors,GoogleAI2021_2,GoAI2021_5,IBM-Berkeley,Utility_3}.

A critical challenge in QErC, specifically leakage of quantum information into non-computational states, is addressed in~\citep{GoogleAI2023_4}. The study introduces a technique using a distance-3 surface code and a distance-21 bit-flip code to actively remove leakage from qubits~\citep{leakage_25,leakage_26,GoAI2022_18}. This approach reduces steady-state leakage by tenfold and maintains an average leakage population of less than $1\times 10^{-3}$, preventing leakage from causing correlated errors and advancing the feasibility of scalable QErC~\citep{GoogleAI2023_1}.

A control optimization strategy~\citep{Thesnakeoptimizer} for managing the frequency trajectories of 68 frequency-tunable~\citep{yan18} superconducting transmon qubits~\citep{GoogleAI2023_1} is presented in~\citep{GoogleAI2023_5}. This strategy improves gate execution and error mitigation, significantly reducing physical error rates compared to non-optimized systems. The approach is projected to be effective for larger systems, such as a distance-23 surface code with 1057 physical qubits, offering a scalable solution adaptable to various quantum algorithms and architectures.

The development of a high dynamic range Josephson parametric amplifier~\citep{Josephsonparametricamplifier}, using an array of rf-SQUIDs, is detailed in~\citep{GoogleAI2023_9}. The amplifier achieves a bandwidth of 250-300 MHz and handles input saturation powers up to -95 dBm at 20 dB gain, tested with a Sycamore (54-qubit) quantum processor. The amplifier exhibits no negative effects on system noise, readout fidelity, or qubit dephasing, with added noise only 1.6 times the quantum limit, addressing gain compression issues common in traditional Josephson parametric amplifiers used for multi-tone multiplexed readout.

The challenges and opportunities in leveraging quantum computing for drug development are examined in~\citep{GoogleAI2023_8}. The perspective highlights the transformative potential of quantum computing in this field and outlines the necessary steps to achieve these advancements~\citep{DrugDiscov,DrugDiscov_AI,DrugDiscov_sycamore}. 
Additionally, the potential for exponential quantum advantage in quantum chemistry is examined in~\citep{GoogleAI2023_3}. The review evaluates the efficiency of quantum state preparation features and classical heuristics~\citep{GoogleAI13,GoogleAI2023_3_6}. The findings suggest that while quantum computers may offer polynomial speedups, exponential speedups for ground-state quantum chemistry are unlikely to be generically achievable.

A review of quantum error mitigation techniques, essential for addressing noise in quantum computers, is provided in~\citep{GoogleAI2023_2}. The review evaluates various mitigation strategies, summarizes current hardware demonstrations, and discusses open challenges. It offers guidance on selecting appropriate techniques based on noise types and explores the potential of these strategies to advance quantum devices in science and industry~\citep{IBM-Berkeley,GoAI2022_11,PennyLane,QREM_0,QREM_1,QuantumMonteCarlo,QuantumLinearSystems,QuantumMetrology,QC_Fermi-Hubbard,Mitiq,optimalentangling,Qermit,QCQiskit,IBMQuantum}. Further details on Google Quantum AI’s research in 2023 can be found in~\citep{GoAI2023_1,GoAI2023_2,GoAI2023_3,GoAI2023_4,GoAI2023_5,GoAI2023_6,GoAI2023_7,GoAI2023_8,GoAI2023_9,GoAI2023_10,GoAI2023_11,GoAI2023_12,GoAI2023_13,GoAI2023_14,GoAI2023_15,GoAI2023_16,GoAI2023_17,GoAI2023_18,GoAI2023_19,GoAI2023_20,GoAI2022_18,GoAI2023_22,GoAI2023_23,GoAI2023_24,GoAI2023_25,GoAI2023_26,GoAI2023_27,GoAI2023_28,GoAI2023_29}.

\subsection{2024: Continuing the quantum quest}

In 2024, the study in~\citep{GoogleAI2023_6} introduces a model-based optimization technique aimed at enhancing the accuracy of quantum measurements in superconducting qubits. This technique achieves a measurement error rate of 1.5\% per qubit with a 500 ns end-to-end duration by addressing issues like excess reset errors and state transitions~\citep{GoAI2022_18,GoogleAI2023_5}. Effective across 17 qubits, this method promises improvements in QErC and other near-term quantum applications.

The research~\citep{GoogleAI2024_1} demonstrates how engineered dissipative reservoirs can direct many-body quantum systems towards useful steady states, with applications in simulating phenomena such as high-temperature superconductivity. Using up to 49-qubit superconducting (transmon) quantum processor, the researchers observed significant results~\citep{GoogleAI2023_1}; including long-range quantum correlations in one dimension and mutual information beyond nearest neighbors in two dimensions. Their findings suggest that engineered dissipation~\citep{Dissipation_8,Dissipation_9,Dissipation_10,Dissipation_11,Dissipation_12,Algorithms,Dissipation_13} could be a scalable alternative to unitary evolution for preparing entangled many-body states on NISQ processors.

In~\citep{GoogleAI2024_2}, a quantum simulator with 69 superconducting qubits is presented, capable of both universal quantum gates and high-fidelity analog evolution. This simulator surpasses classical simulation limits in XEB~\citep{GoogleAI30,GoogleAI27,art19,XEB18} and models a 2D XY quantum magnet. The study reveals signatures of the Kosterlitz-Thouless phase transition~\citep{Kosterlitz-Thouless_3} and deviations from Kibble-Zurek scaling~\citep{Kibble-Zurek_4} due to quantum-classical domain coarsening~\citep{kibble-zurek_5}. It also examines the eigenstate thermalization hypothesis (ETH)~\citep{ETH_6,ETH_7,ETH_8} and investigates energy and vorticity transport in pairwise-entangled dimer states, highlighting the simulator’s effectiveness for exploring many-body spectra and thermalization dynamics~\citep{Thermalizationdynamics_27}.

The method introduced in~\citep{GoogleAI2024_3} uses quantum information theory to simplify many-body Hamiltonians for near-term quantum devices~\citep{similarity_transf_31,similarity_transf_47,similarity_transf_48,similarity_transf_21}. By applying similarity transformations~\citep{similarity_transf_33,similarity_transf_8,similarity_transf_44,similarity_transf_4,AbuGMSc19} as a preprocessing step, this method reduces circuit depth and improves performance on quantum hardware~\citep{similarity_transf_33,similarity_transf_8,similarity_transf_44,similarity_transf_4,similarity_transf_22,similarity_transf_24,similarity_transf_46,similarity_transf_13}. It enhances zero and finite temperature free energy calculations, demonstrating increased effectiveness with higher-quality transformations. This approach represents a practical advance for quantum chemistry on current hardware~\citep{
qchemistry_12,qchemistry_16,qchemistry_7,qchemistry_39,qchemistry_32}.

The work in~\citep{GoogleAI2024_4} explores the interplay between measurement-induced dynamics and conditional unitary evolution, focusing on their effects on entanglement negativity in quantum systems. By analyzing random measurement and feed-forward (MFF) processes~\citep{MFF_29,MFF_33}, the study identifies a sharp transition in the ability to generate entanglement negativity as the number of MFF channels changes. The research links these findings to transitions caused by random dephasing with broken time-reversal symmetry~\citep{time-reversal_24} and rigorously proves the transition using free probability theory~\citep{Measurement-induced phase transitions_25}, with implications for dynamic circuit representations tested on current quantum computing platforms. For further details on Google Quantum AI’s research in 2024, refer to~\citep{GoAI2024_1,GoAI2024_2,GoAI2024_3,GoAI2024_4,GoAI2024_5}.

Through these pioneering efforts, Google Quantum AI continues to pushing the boundaries of what is possible and paving the way for groundbreaking discoveries and applications in quantum computation.

\section{Comparing NISQ processors and classical computers}\label{sec:NISQprocessorsVsclassical}

Currently, NISQ processors are limited to executing a few thousand quantum operations or gates before noise significantly impacts the quantum state~\citep{NISQ18}. Google Quantum AI anticipates that even within this intermediate, noisy regime, their quantum processors will enable applications where quantum experiments can be performed significantly faster than classical supercomputers allow. 

To assess the performance of error-corrected quantum algorithms versus classical algorithms, one can compare their computational costs, as discussed in the field of computational complexity~\citep{computationalcomplexitytheory_1,computationalcomplexitytheory_2,computationalcomplexitytheory_3,computationalcomplexitytheory_4,computationalcomplexitytheory_5,computationalcomplexitytheory_6}. However, this comparison is less straightforward with current experimental quantum processors~\citep{comparenoisy}.

In~\citep{effectiveQV}, Google Quantum AI introduced a framework known as the ``effective quantum volume" (EFQV) for evaluating the computational cost of quantum experiments. This measure quantifies the number of quantum operations or gates involved in producing a measurement outcome. The scheme has been applied to assess the computational cost of various recent experiments, including RCS~\citep{Google_superm_2023}, measurements of ``out-of-time-order correlators" (OTOCs)\citep{OTOCs}, and a recent IBM and UC Berkeley experiment ``Evidence for the utility of quantum computing before fault tolerance," which explores Floquet evolutions related to the Ising model~\citep{IBM-Berkeley}. OTOCs are particularly noteworthy as they offer a direct method for experimentally measuring the EFQV of a circuit, a task that remains computationally challenging for classical computers. Furthermore, OTOCs are valuable in fields such as nuclear magnetic resonance (NMR) and electron spin resonance spectroscopy. Consequently, Google Quantum AI considers OTOC experiments to be promising candidates for demonstrating the practical applications of quantum processors, as detailed in   Section~\ref{OTOCs}.

\subsection{Evaluating the computational cost of noisy quantum processing experiments}

Running a quantum circuit on a ``noisy quantum processor" involves balancing two key factors. On one hand, 
Google Quantum AI aims to tackle problems that are challenging for classical computers~\citep{qsuperm1,nisqQC10}. The computational cost, or the number of operations required for a classical computer to complete a task, is related to the circuit’s EFQV. A larger effective volume generally implies a higher computational cost, indicating that a quantum processor could outperform classical counterparts.

Conversely, each quantum gate on a noisy processor can introduce errors~\citep{QErrorCorrection1,QErrorCorrection2,GoogleAI2023_2}, which degrade the fidelity of the quantum circuit's measurements~\citep{AbuGMSc19,Entangling,SQSCZ2}. As the number of operations increases, so does the error rate, reducing the accuracy of the results~\citep{SycamoreEPJQ,Googles}. Thus, simpler circuits with smaller effective volumes might be preferred, although they are more easily simulated by classical computers. Google Quantum AI seeks to optimize this trade-off, aiming to maximize the ``computational resource," a measure of how well a quantum processor balances these considerations.

A prime example of this trade-off is the RCS experiment~\citep{Google_superm_2023}, a fundamental benchmark in quantum computing. RCS~\citep{GoogleAI30,RCS-5,RCS-31,RCS-BFNV18,RCS-Mov18,RCS2019,RCS-11,GoSup23_21,GoSup23_19,RCS2023,RCSComp2023}, which first demonstrated a quantum processor's ability to surpass a classical computer, is highly sensitive to errors. Any gate error~\citep{QErrorCorrection1,QErrorCorrection2,GoogleAI2023_2} can compromise the experiment, making it a rigorous test of system fidelity while also representing the highest computational cost achievable by a quantum processor. Google Quantum AI recently reported their most advanced RCS experiment to date, featuring a low experimental fidelity of \(1.7 \times 10^{-3}\) and a theoretical computational cost of approximately \(10^{23}\). This experiment, which involved 700 two-qubit gates, would take around 47 years to simulate on the world’s largest supercomputer (at the time of estimation). While this confirms the quantum processor's superiority over classical computers~\citep{comparenoisy,
benchmarkingzeronoise}, it does not represent a particularly practical application on its own.

\subsection{Random circuit sampling}\label{RCS}

Achieving quantum advantage~\citep{qsuperm1,nisqQC10,NISQ24} with noisy processors is a key goal in NISQ computing era~\citep{NISQ18}. A leading approach to this challenge is RCS~\cite{RCS2019}, which involves sampling from probability distributions created by randomly chosen quantum circuits.
Random quantum circuits 
are generally considered challenging for classical systems to simulate~\cite{GoAI2023_1,RCS2023, RCSComp2023}. This challenge has been pivotal in the quest for quantum supremacy~\citep{qsuperm1}, where a quantum computer achieves a task deemed infeasible for classical computers~\citep{qsuperm1,nisqQC10,NISQ18}. 

The aim of RCS~\citep{GoogleAI30,RCS-5,RCS-31,RCS-BFNV18,RCS-Mov18,RCS2019,RCS-11,GoSup23_21,GoSup23_19,RCS2023,RCSComp2023} is to generate a bit string from a distribution that approximates the output distribution of a random quantum circuit. It is important to recognize that a certain level of error is inevitable, as even fault-tolerant quantum computers can only perform RCS with a small acceptable error margin.

From an experimental standpoint, performing RCS on noisy quantum computers without error correction is relatively straightforward. The quality of the samples depends on the fidelity of the quantum device~\cite{GoAI2018_6,GoogleAI30,art19}. Theoretical studies have furthered our understanding of RCS's computational complexity~\cite{RCS-BFNV18,RCS-Mov18}. The RCS supremacy conjecture, in particular, suggests that no classical algorithm can efficiently perform RCS within polynomial time while maintaining a small, reasonable error margin.

Historically, the exploration of circuit sampling began with studying quantum circuits from specific families with structured properties. For instance, simulating instantaneous quantum polynomial (IQP) circuits classically has been notably difficult under reasonable average-case assumptions~\cite{RCS-5,RCS-31}. RCS is also acknowledged as computationally demanding for fully random quantum circuits, according to widely accepted conjectures~\cite{RCS2019,RCS-11,GoAI2023_1}.

RCS has proven to meet average-case hardness conditions~\cite{RCS-BFNV18,RCS-Mov18,RCS-5}, which is crucial for understanding computational difficulty in the presence of experimental noise~\cite{RCS2019}. Additionally, RCS demonstrates an anti-concentration property, meaning errors in estimating output probabilities are relatively small compared to the probabilities themselves~\cite{GoAI2023_1}. These features make RCS a leading candidate for achieving quantum advantage and drive its use in experiments aimed at demonstrating quantum supremacy~\cite{RCS2019}. For a more detailed examination of RCS and its computational complexity, readers are referred to~\cite{GoAI2023_1,RCS2023, RCSComp2023}.

\subsection{The OTOCs and Floquet evolution}\label{OTOCs}

Many complex questions in quantum many-body physics~\citep{many-bodyQuestions1} remain beyond classical computation, presenting significant opportunities for quantum processors~\citep{many-bodyQuestions2}. Unlike RCS, which measures the entire quantum state at the end of an experiment, these investigations often focus on specific local observables. The EFQV for a local observable can be smaller than that of the complete circuit due to the localized impact of quantum operations on the observable~\citep{effectiveQV}.

\begin{figure}
    \centering
    \includegraphics[width=0.37\textwidth]{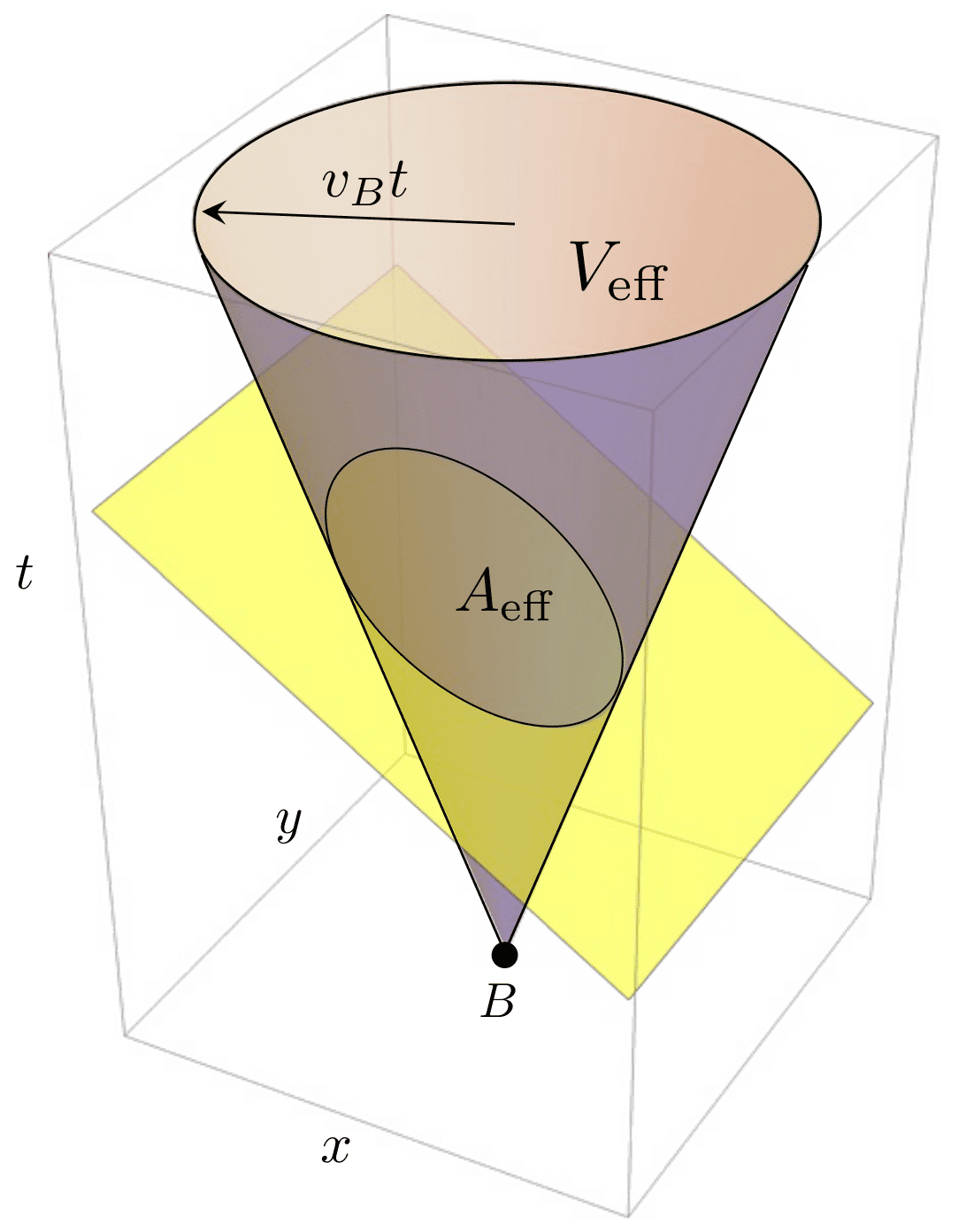}
\caption{ A schematic representation shows a conical surface in the ($x, y, t$) space, encompassing the effective volume \( V_{\text{eff}} \) that envelops the tensorial structure of entangling gates affecting the local observable \( B(t) = U^\dagger(t) B U(t) \), with \( B(0) = B \) indicated by a black dot. The cross-section of this cone, with area \( A_{\text{eff}} \), is highlighted where the plane intersects the cone. The cone’s base represents the subset of qubits in the ($x, y$) plane engaged in the operator spreading of \( B \) at time \( t \), while the perimeter of this base signifies the scrambling front advancing with velocity \( v_B \). 
Reproduced under a creative common license (\url{ https://creativecommons.org/licenses/by/4.0/}) from~\citep{effectiveQV}.}
    \label{fig:cone}
\end{figure}

To illustrate this, Google Quantum AI use the concept of a ``\textit{butterfly cone}," analogous to the light cone in the ``theory of relativity," which defines the causal connections between events. In quantum systems, the \textit{butterfly cone} (depicted in  Figure~\ref{fig:cone}) represents the area where quantum information spreads, with its growth rate determined by the butterfly speed—measured by OTOCs. The EFQV of a local observable corresponds to the volume within this \textit{butterfly cone}, encompassing only those operations causally related to the observable. Thus, faster information spreading results in a larger effective volume, making classical simulation more challenging.

In~\citep{effectiveQV}, Google Quantum AI explores and elucidates the trade-off between the sensitivity of an experimental observable to noise and the corresponding classical computational cost required to evaluate this observable. Given an operator ${\cal Q}$ with an ideal expectation value $\langle {\cal Q} \rangle_{\text{ideal}}$ and an experimental expectation $\text{tr}(\pounds {\cal Q})$, where $\pounds$ denotes a noisy quantum state density matrix. Their findings suggest that, under certain experimental conditions~\citep{effectiveQV}, we have:
\begin{equation}
\text{tr}(\pounds {\cal Q}) = {\cal F}_{\text{eff}} \langle {\cal Q} \rangle_{\text{ideal}},
\end{equation}

\noindent
where ${\cal F}_{\text{eff}}$ represents the effective fidelity of the observable. This fidelity is anticipated to decrease exponentially as:
\begin{equation}
{\cal F}_{\text{eff}} \sim e^{-\varpi V_{\text{eff}}},
\end{equation}
\noindent
with $V_{\text{eff}}$ the effective circuit volume and $\varpi$ being the dominant error per 2-qubit entangling gate. For systems lacking conservation laws, $V_{\text{eff}}$ equates to the number of entangling 2-qubit gates influencing the expectation value $\langle {\cal Q} \rangle_{\text{ideal}}$. The computational cost associated with tensor network contraction is expected to grow exponentially with an effective area $A_{\text{eff}}$ related to a specific cut of the effective volume $V_{\text{eff}}$, given by:
\begin{equation}
\text{cost} \propto 2^{\gamma A_{\text{eff}}},
\end{equation}
\noindent
where $\gamma$ is a constant. Consequently, Google Quantum AI anticipates a trade-off between achieving a high signal-to-noise ratio and incurring high classical computational costs (large $A_{\text{eff}}$). A depiction of the EFQV, \( V_{\text{eff}} \), associated with the gates influencing the local observable \( B \) is shown in Figure~\ref{fig:cone}. Additionally, the effective area, \( A_{\text{eff}} \), is illustrated as the cross-sectional area of the cone. The perimeter of the cone’s base represents the boundary of information propagation, moving at the butterfly velocity, \( v_B \). For further details on this framework, especially in the context of RCS where it is more naturally applicable, see~\citep{effectiveQV}.

In a recent experiment involving a Floquet Ising model~\citep{IBM-Berkeley}, related to time crystal and Majorana edge mode studies, a pioneering experiment by IBM Quantum and UC Berkeley unveiled a significant advancement towards practical quantum computing~\citep{IBM-Berkeley}. The study revealed that quantum computers can perform circuits that surpass the limits of classical brute-force simulations. Notably, IBM Quantum~\citep{IBMQuantum} now possesses both hardware and software capable of running quantum circuits involving 100 qubits and 3,000 gates, all without prior outcome knowledge~\citep{Utility_3}.

Later on, Google quantum AI applied its scheme~\citep{effectiveQV} to the data from~\citep{IBM-Berkeley}. They noted that an effective fidelity of 0.37 was estimated for the largest circuits. Given a gate error rate of approximately 1\%, this corresponds to an effective volume of around 100, significantly smaller than the light cone volume, which includes about 2,000 gates across 127 qubits. Numerical simulations suggest that the effective volume is closer to 28 qubits. This discrepancy~\citep{benchmarkingzeronoise} in effective volume is consistent with OTOC measurements. Despite the experiment's depth, the computational cost is about \(5 \times 10^{11}\), nearly a \textit{trillion} times less than that of the recent RCS experiment. Consequently, this experiment can be simulated in under a second per data point on an A100 GPU, making it a valuable application but not yet surpassing classical simulations~\citep{effectiveQV}.

OTOC experiments hold promise for computational applications by revealing critical physical insights~\citep{OTOCs}, such as butterfly velocity, which is essential for accurately measuring a circuit's EFQV. Experiments with rapid entangling gates may pave the way for the first beyond-classical demonstration using quantum processors. Google quantum AI's 2021 OTOC experiment~\citep{OTOCs} achieved an effective fidelity of approximately 0.06 with a signal-to-noise ratio of $\approx 1$, resulting in an effective volume of around 250 gates and a computational cost of \(2 \times 10^{12}\).

Although current OTOC experiments do not yet exceed classical simulation capabilities, their potential is significant. Quantum circuits exploring extensive energy levels, which are typically chaotic, are hard to simulate classically and quickly decay in standard time-order correlators (TOCs)~\citep{Gocircuit1}. OTOCs, however, allow for growing complexity, limited primarily by gate errors. A reduction in error rates could double the computational cost, moving the experiment into the beyond-classical regime.

\section{The Quantum Computational Supremacy} \label{sec:Supremacy}

In this section, we explore Google Quantum AI's pursuit of quantum computational supremacy~\citep{qsuperm1,nisqQC10,NISQ18}. For a comprehensive review of the broader quest for quantum supremacy, including experiments beyond those conducted by Google, as well as a variety of pioneering proof-of-principle experiments in various quantum computing domains, readers are encouraged to consult~\citep{NISQ24}.

\subsection{The XEB theory}

To confirm the accurate performance of the Google's quantum processors during the quantum supremacy experiment, Google Quantum AI and its collaborators utilize a technique known as cross-entropy benchmarking (XEB)~\citep{GoogleAI30,RCS2019,GoogleAI27}. XEB evaluates how closely the distribution of observed bit strings from the quantum processor matches the expected probabilities from classical simulations~\citep{GoogleAI30,GoogleAI27,art19,XEB18}. This method allows for a rigorous performance assessment by comparing experimental results with theoretical predictions~\citep{art19}.

XEB~\citep{GoogleAI27,art19-25} offers a method for calibrating single- and 2-qubit gates and assessing fidelity in RQCs~\citep{GoogleAI30,RCS-5,RCS-31,RCS-BFNV18,RCS-Mov18,RCS2019,RCS-11,GoSup23_21,GoSup23_19,RCS2023,RCSComp2023}, even those with numerous qubits. It capitalizes on the analogy between the measurement probabilities of random quantum states and the patterns of laser ``speckles," where some bit strings are more likely than others~\citep{art19-27,art19-28}. 
By analyzing distortions in these speckle patterns, XEB can assess error rates and fidelity without needing to reconstruct complex experimental output probabilities, which would be impractical due to the exponential increase in measurements with more qubits~\citep{art19,GoogleAI30,art19-13,GoogleAI27}.

The XEB fidelity ($\mathcal{F}_{\text{XEB}}$) gauges how often high-probability bit strings are sampled, 
with values ranging from 0 to 1 reflecting the degree of error in the circuit execution. Google Quantum AI aims to achieve a high $\mathcal{F}_{\text{XEB}}$ for a quantum circuit with adequate width and depth, making classical computation costs prohibitive. This goal is challenging due to the imperfections in logic gates and the susceptibility of the quantum states to errors~\citep{art19}. Even a small error, such as a bit or phase flip during the algorithm's execution, can drastically alter the speckle pattern, resulting in nearly \textit{zero} fidelity~\citep{GoogleAI30}. Therefore, quantum processors capable of executing circuits with minimal error rates are required to demonstrate quantum supremacy~\citep{Google_superm_2023}.

\subsection{Quantum processors}

\begin{figure*}
    \centering
    \includegraphics[width=0.9 \textwidth]{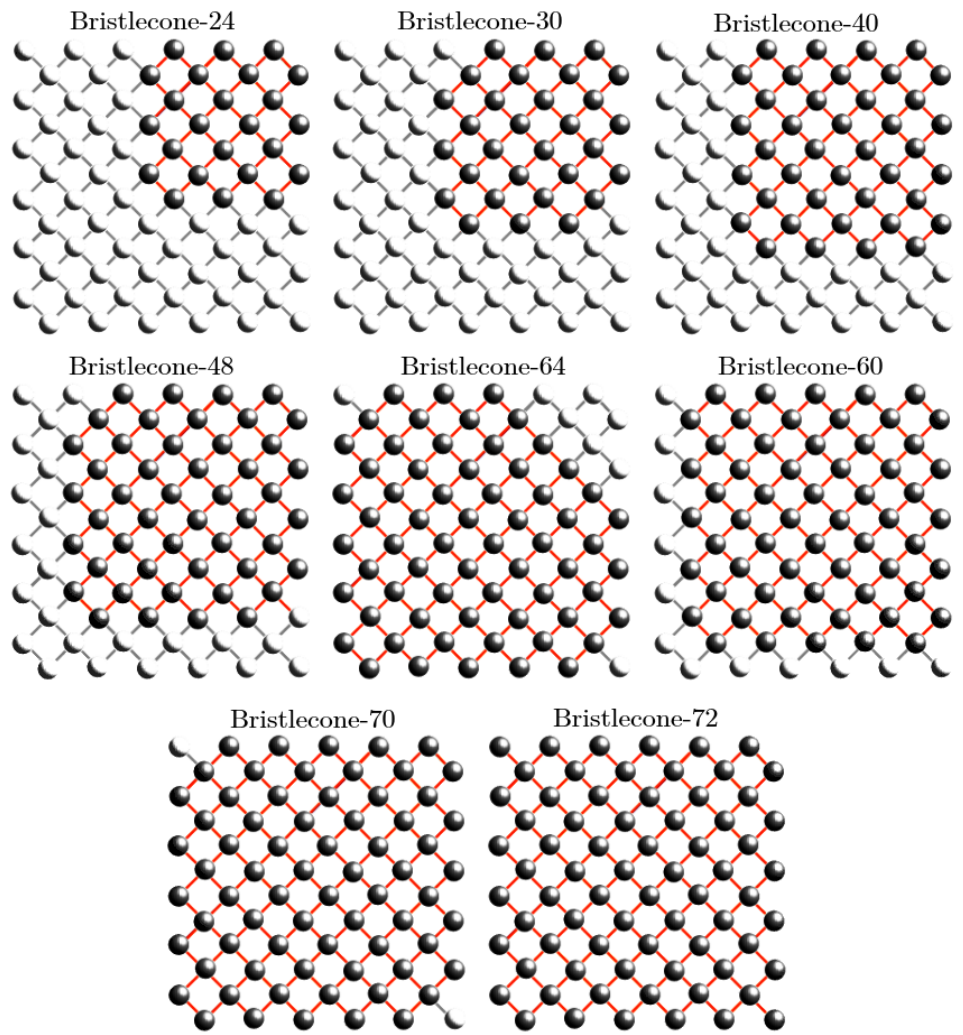}
    \caption{Architectural overview of Google's \textit{Bristlecone} quantum processor. The sub-lattices of the complete \textit{Bristlecone}-72 (shown in the bottom right) are arranged in order of increasing complexity for a given depth. Notably, \textit{Bristlecone}-72 is not more difficult to simulate than \textit{Bristlecone}-70, as the corner tensors can be contracted with minimal computational expense. Additionally, \textit{Bristlecone}-64 exhibits similar simulation complexity to \textit{Bristlecone}-48, while being significantly easier to simulate than \textit{Bristlecone}-60~\citep{GoogleAI36}. Google Quantum AI has identified a series of sub-lattices, specifically \textit{Bristlecone}-24, \textit{Bristlecone}-30, \textit{Bristlecone}-40, \textit{Bristlecone}-48, \textit{Bristlecone}-60, \textit{Bristlecone}-64, and \textit{Bristlecone}-70, arranged from top left to bottom left, all of which present significant challenges for classical simulation while maintaining a low qubit count.}
    \label{fig:Bristlecone}
\end{figure*}

Google Quantum AI's quantum processors are at the cutting edge of computational technology, designed to address problems that exceed the capabilities of classical computers~\citep{art19,Google_superm_2023}. The company has developed several significant quantum processors, including \textit{Foxtail}~\citep{9array1,9array2}, \textit{Bristlecone}~\citep{Bristlecone}, and \textit{Sycamore}~\citep{art19}. These processors are characterized by their advanced control and precision, enabling the manipulation and measurement of qubits with remarkable accuracy~\citep{art19,Google_superm_2023}.

\textit{Sycamore:} The \textit{Sycamore} processor, a key player in Google Quantum AI's portfolio, utilizes superconducting qubits and gained widespread attention for its role in achieving quantum supremacy in 2019~\citep{art19}. The SYC-53, specifically, is engineered to perform intricate quantum computations using a 2-dimensional array of 54 transmon qubits~\citep{koch07,Stef14}. It stands out for its ability to execute high-fidelity 1-qubit and 2-qubit quantum gates, which are crucial for conducting practical and complex quantum operations~\citep{SycamoreEPJQ,Googles}.

\textit{Bristlecone:} \textit{Bristlecone} QPUs~\citep{GoogleAI36} represent a significant advancement in quantum processor design with a larger array of qubits compared to \textit{Sycamore}. This processor aims to enhance the scalability and performance of quantum computing~\citep{GoogleAI36}. \textit{Bristlecone's} design focuses on improving qubit coherence times and reducing error rates, setting the stage for more sophisticated quantum computations and demonstrating the progress towards more robust and scalable quantum systems. An architectural overview of Google’s \textit{Bristlecone}-72 quantum processor and series of sub-lattices (\textit{Bristlecone}-24, \textit{Bristlecone}-30, \textit{Bristlecone}-40, \textit{Bristlecone}-48, \textit{Bristlecone}-60, \textit{Bristlecone}-64, and \textit{Bristlecone}-70) are shown in Figure~\ref{fig:Bristlecone}.

\begin{figure}
    \centering
    \includegraphics[width=0.45 \textwidth]{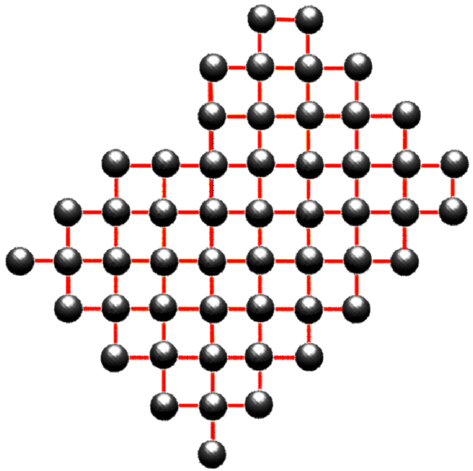}
    \caption{Qubit grid of the \textit{Weber} quantum computer (part of the \textit{Sycamore} family), a universal gate-based quantum computer, features 53 superconducting qubits.}
    \label{fig:Weber}
\end{figure}

\textit{Weber:} Google's universal gate-based superconducting quantum computer, known as \textit{Weber} (part of the \textit{Sycamore} family). The qubit grid of weber is shown in Figure~\ref{fig:Weber}, general specifications are detailed in Table~\ref{t:google}. Although these values reflect historical data~\citep{Weber}, they are important for documenting the evolution of Google's quantum computing technology in the NISQ era~\citep{NISQ18}.

Recent performance metrics, including system characteristics of Google's \textit{Sycamore} quantum processors during the quantum supremacy experiments, are discussed in the subsequent sections~\ref{subsec:SYC53} and~\ref{subsec:SYC67SYC70}. 
This section provides updated insights into the hardware performance of these advanced quantum processors, highlighting their capabilities and improvements achieved over time. These quantum processors exemplify Google Quantum AI's dedication to pushing the frontiers of quantum computing.

\begin{figure}
    \centering
    \includegraphics[width=0.4 \textwidth]{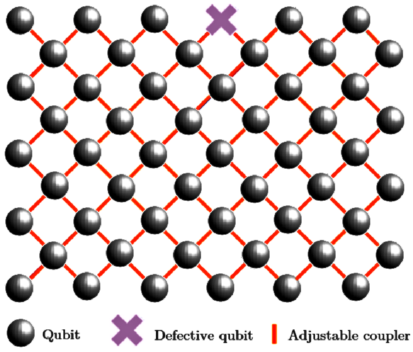}
    \caption{Architecture of Google's Sycamore (SYC-53) quantum processor. 
    The processor features a rectangular array of 54 programmable superconducting transmon qubits, with adjustable couplers linking each qubit to its four nearest neighbors. Due to one non-functional qubit during the supremacy regime experiment~\citep{art19}, the device operated with 53 qubits and 86 couplers.}
    \label{fig:Sycamore}
\end{figure}

\begin{figure*}
    \centering
    \includegraphics[width=\textwidth]{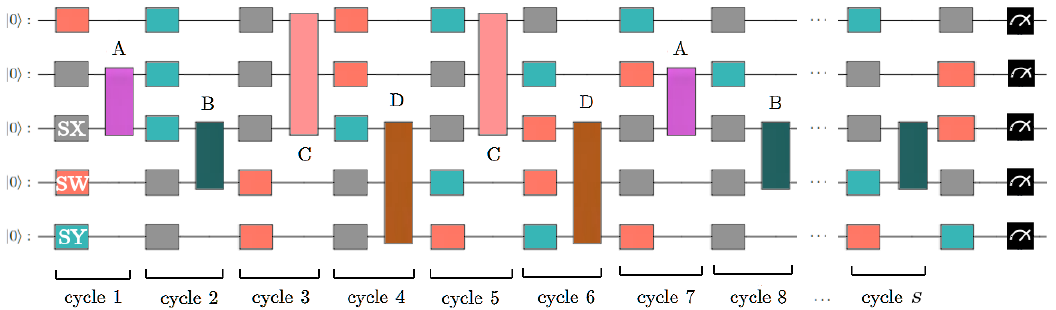}
    \caption{Schematic diagram of an $s$-cycle circuit for the SYC-53 RQCs. Each cycle includes a layer of random single-qubit gates (selected from $\{X^{1/2}, W^{1/2}, Y^{1/2} \}$), followed by a layer of 2-qubit gates labeled A, B, C, or D. In longer circuits, the sequence of layers repeats as A; B; C; D -- C; D; A; B. 1-qubit gates are not repeated consecutively, and there is an additional layer of 1-qubit gates before measurement.}
    \label{fig:RQCs}
\end{figure*}

\begin{table*}
\caption{\label{t:google}The general service specifications of Google's universal gate-based superconducting quantum processor, \textit{Weber} (\textit{Sycamore} family)~\citep{Weber}. Recorded for historical documentation in the NISQ era literature.} 
\centering
    \begin{tabular}{c}
    \includegraphics[width=0.95\textwidth]{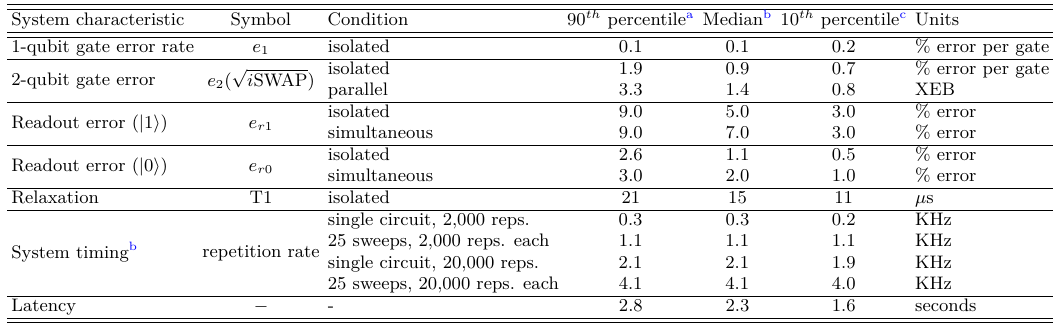} \\
    \end{tabular}
\footnotetext[1]{The 90th percentile value from the overall distribution of post-calibration characterizations for all qubits from January to March 2021.}
\footnotetext[2]{The median value derived from the 90-day distribution of median values across all qubits from January to March 2021.}
\footnotetext[3]{The 10th percentile value from the aggregated distribution of post-calibration characterizations for all qubits between January and March 2021.}
\end{table*}

\begin{table*}
\caption{\label{t:ggp} Aggregate system characteristics of Google's SYC-53 quantum processor~\citep{art19}.} 
\centering
    \begin{tabular}{c}
    \includegraphics[width=0.72\textwidth]{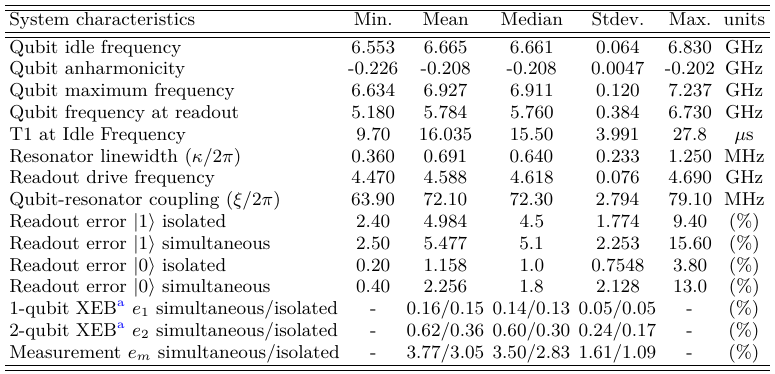} \\
    \end{tabular}
\end{table*}

\begin{table*}
\caption{\label{table:cost}
Estimated computational costs for simulations with various superconducting quantum computers: SYC-53, SYC-67, SYC-70, Zuchongzhi 2.0, and Zuchongzhi 2.1. 
The fifth column shows the number of FLOPs required to compute a single output amplitude without memory constraints. 
The subsequent three columns detail the computational resources required to simulate noisy sampling of one million bit-strings. 
These estimates are based on \textit{Frontier}, a leading supercomputer with a theoretical peak performance of $1.685 \times 10^{18}$ single-precision FLOPS. A 20\% efficiency in FLOP utilization is assumed~\citep{GooSup23-14,Cupjin,feng22}, and the simulation’s lower fidelity is factored into the cost~\citep{GooSup23-14,Cupjin,GoAI2018_6,GoogleAI36}. Each single-precision complex FLOP requires eight machine FLOPs. For SYC-67, costs are estimated assuming memory is distributed across all RAM$^{\models}$ or secondary storage$^{\notin}$, without considering bandwidth constraints. Other entries use tensor contraction algorithms that are easily parallelizable across GPUs~\citep{Cupjin,feng22,GooSup23-31}. Adapted under a Creative Commons license (\url{https://creativecommons.org/licenses/by/4.0/}) from~\citep{Google_superm_2023}.}
\centering
    \begin{tabular}{c}
    \includegraphics[width=0.95\textwidth]{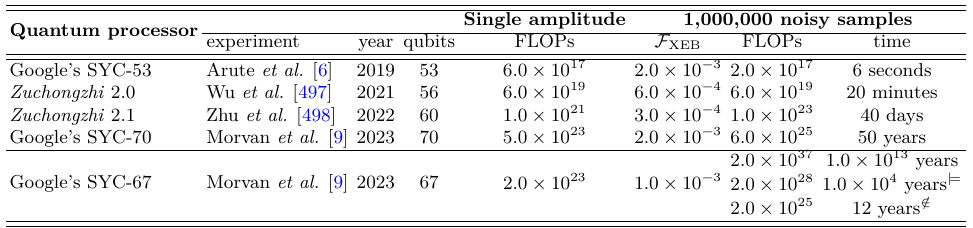} \\
    \end{tabular}
\end{table*}

\subsection{The 53-qubits \textit{Sycamore} processor}\label{subsec:SYC53}

Google's \textit{Sycamore} (SYC-53) quantum processor~\citep{art19} features a two-dimensional grid of 54 superconducting transmon qubits~\citep{koch07}. Each qubit is linked to its four nearest neighbors, creating a rectangular lattice, as shown in Figure~\ref{fig:Sycamore}. A landmark achievement of the SYC-53 processor was its ability to perform a computation in just 200 seconds that would  have taken a classical computer roughly 10,000 years~\citep{art19}.

In this experiment\citep{art19}, Google's Quantum AI team programmed the SYC-53 quantum processor to perform RCS~\citep{GoogleAI30,RCS-5,RCS-31,RCS-BFNV18,RCS-Mov18,RCS2019,RCS-11,GoSup23_21,GoSup23_19,RCS2023,RCSComp2023}. 
This task was executed in a computational space of $9 \times 10^{15}$—a scale too complex for traditional machines. The initial estimation for performing this computation on \textit{Summit}, the leading supercomputer at the time of Google's quantum supremacy demonstration, was about 10,000 years. In contrast, the SYC-53 completed the same task in just 200 seconds. This milestone demonstrated the potential of quantum computing~\citep{NISQ24} to tackle problems that are beyond the reach of classical methods.

However, after the publication, a discussion emerged about the potential overestimation of the time required to tackle the same problem on a supercomputer. Pednault \textit{et al.}~\citep{Pednault} later argued that a similar task could be achieved with high accuracy using a classical supercomputer like \textit{Summit}, within a few days. Additionally, Huang \textit{et al.}~\citep{Cupjin} presented a classical simulation algorithm based on tensor networks that can accomplish the task in under 20 days. Another tensor network approach, detailed by Pan \textit{et al.}~\citep{feng22}, addressed the sampling problem of Sycamore’s quantum circuits and was executed on a computational cluster with 512 GPUs, completing the computation in 15 hours.

Google Quantum AI and its collaborators tested the largest circuit, which included 53 qubits and 20 cycles. They collected \( 30 \times 10^6 \) samples from ten runs of the circuit, finding an observed $\mathcal{F}_{\text{XEB}}$ of \( (2.24 \pm 0.21) \times 10^{-3} \) for the truncated circuits. They assert that the average fidelity of running these circuits on the quantum processor is at least 0.1\%~\citep{art19}.

To reach this quantum supremacy milestone~\citep{qsuperm1,GoogleAI27}, the SYC-53 processor incorporated several technological innovations, as detailed in~\citep{art19}. These advancements allowed the execution of RQCs composed of alternating layers of 1-qubit and 2-qubit gates, organized into cycles (as depicted in Figure~\ref{fig:RQCs}). Each random circuit includes $s$-cycles, where a cycle is defined by a sequence of one-qubit gates, followed by two-qubit gates, and concluding with another layer of single-qubit gates before measurement. 

The SYC-53 uses transmon qubits~\citep{koch07,Stef14}, which are nonlinear superconducting resonators operating between 5 and 7 GHz. Each qubit utilizes the two lowest energy states of the resonant circuit and has two control mechanisms: a microwave drive for excitation and a magnetic flux control for frequency tuning. For state readout, each qubit is linked to a linear resonator~\citep{wall04}, and adjustable couplers~\citep{chen14,yan18} enable dynamic tuning of qubit coupling from fully disengaged to 40 MHz. During the supremacy experiment, one qubit malfunctioned, leaving the device operational with 53 qubits and 86 couplers~\citep{NISQ18}. For more information on Google's SYC-53 quantum AI, refer to~\citep{SycamoreEPJQ, Googles}, which details the characterization of the \textit{Sycamore} quantum processes through comprehensive quantum tomography experiments.

\subsection{The SYC-67 and SYC-70 quantum processors}\label{subsec:SYC67SYC70}

In late 2023, Google Quantum AI and its collaborators conducted groundbreaking experiments~\citep{Google_superm_2023} using quantum processors with 67 (SYC-67) and 70 (SYC-70) qubits, targeting the realm of quantum supremacy~\citep{qsuperm1,nisqQC10,NISQ24}. These experiments provided valuable insights into how quantum dynamics interact with noise~\citep{Google_superm_2023}. The identified phase boundaries offer crucial information about the conditions needed for noisy quantum devices to maximize their computational potential. Notably, they achieved new results in RCS~\citep{GoogleAI30,RCS-5,RCS-31,RCS-BFNV18,RCS-Mov18,RCS2019,RCS-11,GoSup23_21,GoSup23_19,RCS2023,RCSComp2023} with an estimated fidelity of $1.5 \times 10^{-3}$ for 67 qubits and 32 cycles (or 880 entanglement gates). This represents a significant expansion in circuit complexity compared to earlier studies~\citep{art19}, while maintaining the same fidelity level.

To realistically assess the computational resources needed to simulate RCS~\citep{GoogleAI30,RCS-5,RCS-31,RCS-BFNV18,RCS-Mov18,RCS2019,RCS-11,GoSup23_21,GoSup23_19,RCS2023,RCSComp2023}, one must account for the limitations of supercomputers, including FLOPS, memory capacity, and bandwidth constraints. Table~\ref{table:cost} provides estimates for simulating the largest RCS instances from previous studies~\citep{art19,Zuchongzi,Zuchongzi2.1} and the recent Google study~\citep{Google_superm_2023}. These estimates consider sampling 1 million uncorrelated bit-strings with fidelity comparable to the experimental results, using the top-performing supercomputer, \textit{Frontier}, which has a theoretical peak performance of $1.7 \times 10^{18}$ single-precision FLOPS across GPUs with 128 GB of RAM each.

By demonstrating an RCS experiment with 67 qubits at 32 cycles, Google Quantum AI and its collaborators show that their computational requirements exceed the capabilities of current classical supercomputers, even in the presence of unavoidable noise. Additionally, the dominance of global correlations over XEB~\citep{GoogleAI30,GoogleAI27,art19,XEB18} in the weak noise phase offers protection against ``spoofing" attacks, a notable distinction from Boson-Sampling (BoS) experiments~\citep{BosonS}. Recent BoS metrics~\citep{jiuzhang,Jiuzhang2.0,Borealis} have been predominantly influenced by local correlations~\citep{GoogleAI49}. For a more exploration of BoS and Gaussian Boson-Sampling (GBoS) experiments, as well as their significance in achieving quantum supremacy with photonic quantum computers~\citep{Borealis5}, readers are directed to~\citep{Light,PhotonicQuantumComputers}.

\begin{figure*}
    \centering
    \includegraphics[width=0.87\textwidth]{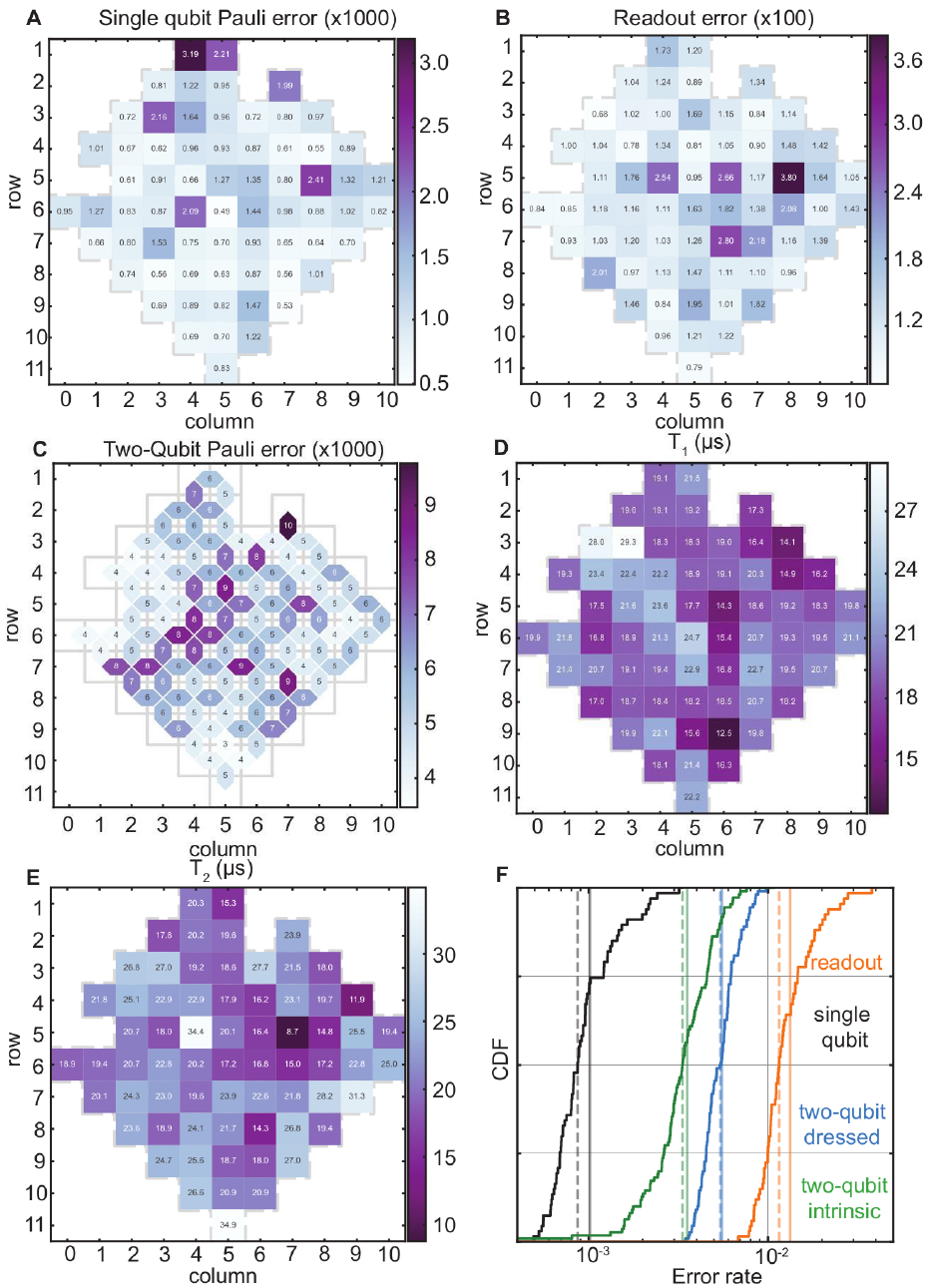}
    \caption{SYC-67 Device Benchmarking: 
    A: Single-qubit Pauli error rate from Randomized Benchmarking. 
    B: Readout error rate averaged from random bitstrings. 
    C: Two-qubit Pauli error rate via parallel 2-qubit XEB. 
    D: T1, and E: T2 echo times. 
    F: Cumulative distribution function (CDF) of errors, with the solid vertical line indicating the average and the dashed line showing the median. Regenerated under a Creative Commons license (\url{https://creativecommons.org/licenses/by/4.0/}) from~\citep{Google_superm_2023}.}
    \label{fig:SYC67}
\end{figure*}

\begin{figure*}
    \centering
    \includegraphics[width=0.87\textwidth]{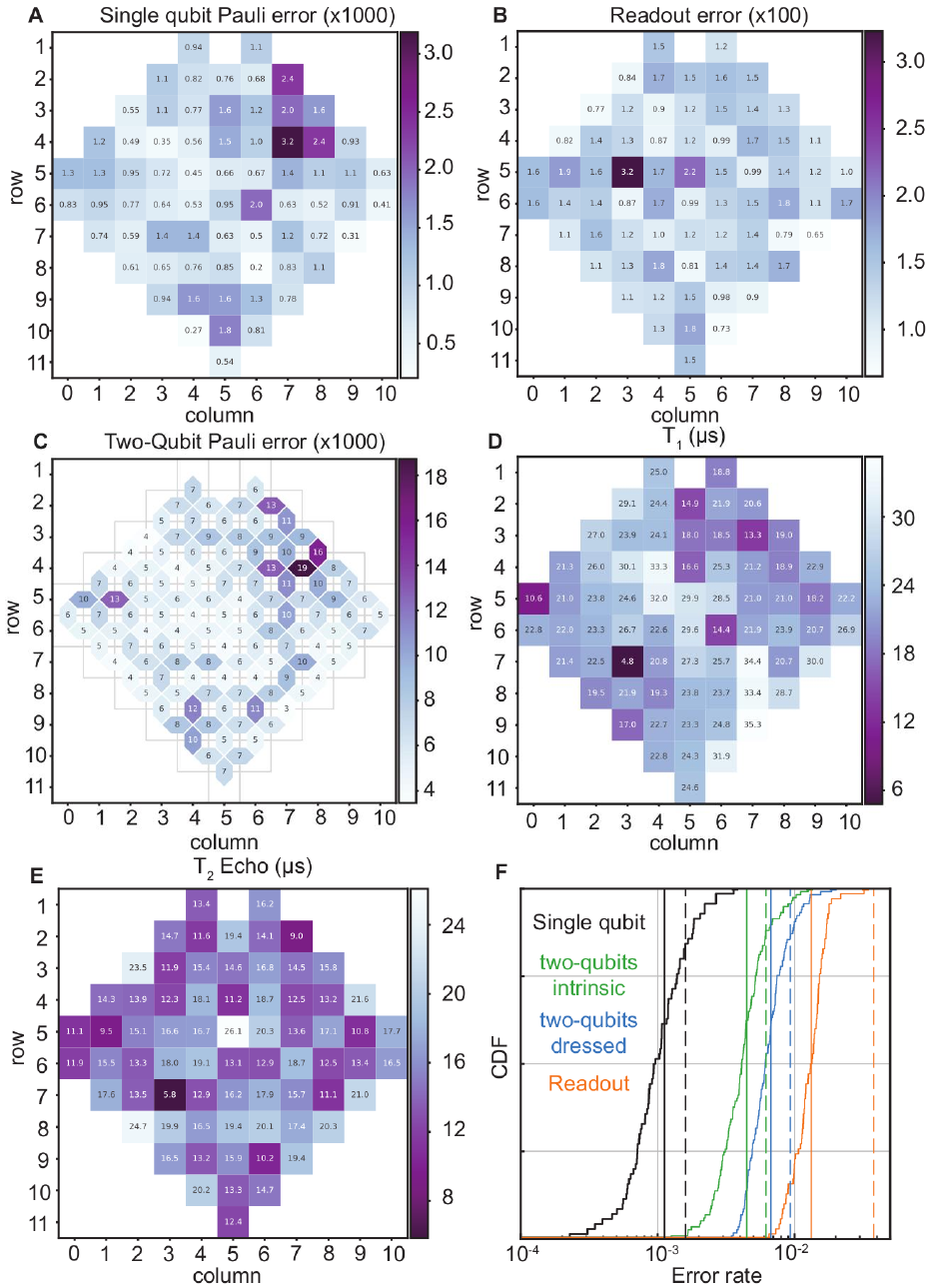}
    \caption{Device Benchmarking for SYC-70: Assessment of random circuit elements. 
A: Single-qubit Pauli error rate via Randomized Benchmarking. 
B: Readout error rate from averaging random bitstrings. 
C: Two-qubit Pauli error rate using parallel 2-qubit XEB. 
D: T1, and E: T2 echo times. 
F: CDF of various error types, with the solid vertical line denoting the average and the dashed line representing the average from Ref.~\citep{art19}. Regenerated under a Creative Commons license (\url{https://creativecommons.org/licenses/by/4.0/}) from~\citep{Google_superm_2023}.}
\label{fig:SYC70}
\end{figure*}

\begin{figure*}
    \centering
    \includegraphics[width=0.95\textwidth]{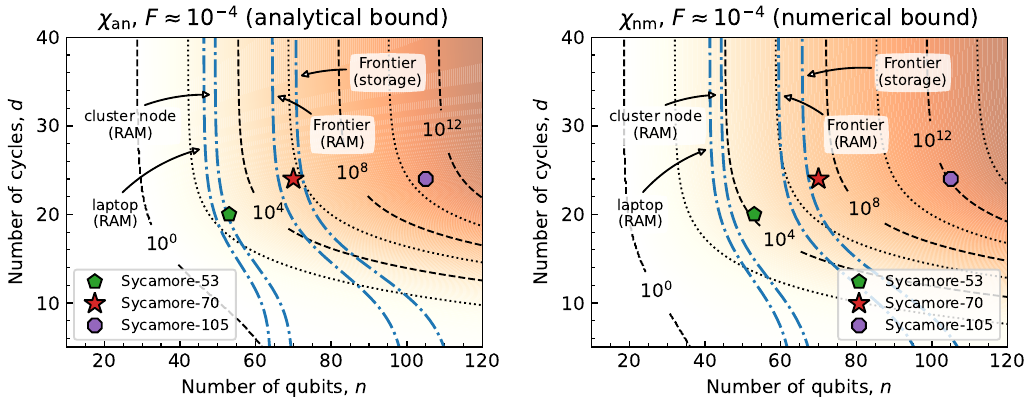}
\caption{Upper bounds on the required bond dimension for achieving a target fidelity of ${\cal F} = 10^{-4}$, shown analytically (left) and numerically (right) as functions of qubit count and circuit depth. The memory footprint (blue dashed-and-dotted lines) represents the memory needed to store two complex 32 tensors of dimension $2^{n/2} \times \chi$. Memory specifications are: Laptop (RAM) = 32GB, cluster node (RAM) = 256GB, Frontier (RAM) = 9.2PB, and Frontier (storage) = 700PB. The density map averages over circuits with patterns A;B;C;D;C;D;A;B/B;A;D;C;D;C;B;A and diagonal/vertical cuts. Regenerated under a Creative Commons license (\url{https://creativecommons.org/licenses/by/4.0/}) from~\citep{Google_superm_2023}.}
    \label{fig:ChivsChi}
\end{figure*}

Figures~\ref{fig:SYC67} and \ref{fig:SYC70} offer critical insights into the performance of the SYC-67 and SYC-70 devices, respectively. Both figures detail the performance metrics and benchmarking results, including 1-qubit and 2-qubit Pauli error rates, readout error rates, and T1 and T2 echo times, along with the cumulative distribution function (CDF) of various errors~\citep{Google_superm_2023}.

Figure~\ref{fig:ChivsChi} illustrates the upper bounds on the required bond dimension to achieve a target fidelity of ${\cal F} = 10^{-4}$, comparing analytical and numerical results as functions of qubit count and circuit depth. It also details the memory footprint needed to store complex tensors of dimension ($2^{n/2} \times \chi$), with varying memory capacities for different computing resources, including laptops, cluster nodes, and high-performance systems like \textit{Frontier}. Additionally, the figure shows the density map, which is obtained by averaging over circuits with specific patterns, including A;B;C;D;C;D;A;B/B;A;D;C;D;C;B;A and various diagonal or vertical cuts~\citep{Google_superm_2023}. By continually advancing their hardware capabilities, Google Quantum AI is driving the development of quantum technology and facilitating breakthroughs across various fields.

\section{Quantum Software}\label{sec:Software}

Google Quantum AI is advancing the field of quantum computing; creating powerful tools that enable researchers to go beyond the limits of classical computing~\citep{qsuperm1,nisqQC10,NISQ24}. Google's dedicated software and hardware solutions are specifically crafted to facilitate the development of innovative quantum algorithms capable of tackling practical problems in the near term~\citep{NISQAlgorithms}. They have created various tools, which we explore in this section.

\subsection{Cirq}

Cirq is a \textit{Python} software library for creating, manipulating, and optimizing quantum circuits~\citep{Cirq}. It enables precise control over quantum gates and operations, allowing researchers to fully utilize quantum computers and simulators. Developed by Google Quantum AI, this open-source framework addresses the unique challenges of NISQ devices~\citep{NISQ18,NISQ24}. Cirq provides essential tools for working with NISQ systems~\citep{qsuperm1,nisqQC10,NISQ24}, where detailed hardware knowledge is crucial for optimal performance~\citep{QErrorCorrection1,QErrorCorrection2,GoogleAI2023_2}. Additionally, Cirq is specifically tailored for crafting innovative quantum algorithms for near-term quantum computers~\citep{NISQAlgorithms}.

\subsection{OpenFermion}

Another valuable library developed by Google Quantum AI is OpenFermion~\citep{OpenFermion0}. OpenFermion is an open-source tool for translating chemistry and materials science problems into quantum circuits that can be executed on current quantum computing platforms~\citep{qsuperm1,nisqQC10,NISQ24}. It specializes in compiling and analyzing quantum algorithms specifically designed for simulating fermionic systems, including quantum chemistry~\citep{qsimulations_Feynman,qsimulations_Lloyd,qsimulations_2019,qchemistry_12,qchemistry_16,qchemistry_7,qchemistry_39,qchemistry_32}. The library offers efficient data structures for fermionic operators and circuit primitives for execution on quantum devices. Its plugins further streamline the translation of electronic structure calculations into quantum circuits~\citep{OpenFermion}.

Recent research highlights OpenFermion’s application in variational simulations of chemistry. The study in~\citep{OpenFermion} involved simulating molecules with up to a dozen qubits, demonstrating the use of error-mitigation strategies and parameterized ansatz circuits~\citep{GoogleAI2016_1} to prepare Hartree-Fock wavefunctions~\citep{Hartree-Focktheory}. These experiments benchmark the hardware performance and lay the groundwork for more complex simulations of molecular and correlated systems~\citep{qsimulations_Feynman,qsimulations_Lloyd,qsimulations_2019,qchemistry_12,qchemistry_16,qchemistry_7,qchemistry_39,qchemistry_32}, illustrating OpenFermion’s utility in advancing quantum chemistry and material science research.

\subsection{TensorFlow Quantum}

TensorFlow Quantum~\citep{TensorFlow_whitepaper} is a unique library that combines the power of quantum computing with classical machine learning. It enables researchers to rapidly prototype hybrid quantum-classical machine learning models, opening up new avenues for exploration and innovation~\citep{drugdiscovery,Quantumcomputationalchemistry,Qalgorithms-chemistry-materials,quantumbiology,HybridML-drugdesign}.

It facilitates the design and training of both discriminative and generative quantum models while leveraging high-performance quantum circuit simulators. TFQ supports seamless integration with quantum algorithms and logic from Google’s Cirq~\citep{Cirq}, providing essential quantum computing primitives that are compatible with TensorFlow’s existing APIs. This framework is designed for exploring and developing models for quantum data and applications, supporting a range of applications, from basic supervised learning and quantum control to advanced quantum learning techniques~\citep{learningtolearn_38,learningtolearn_35,learningtolearn_34,learningtolearn_36,learningtolearn_37}. For further details, refer to~\citep{TensorFlow_whitepaper}.

\subsection{Qsim}

Qsim is an advanced (optimized) quantum circuit simulator that seamlessly integrates with Cirq~\citep{Cirq}, offering high-performance capabilities. It is a comprehensive wave function simulator developed in C++. Developed in C++, it serves as a comprehensive wave function simulator. Qsim utilizes AVX/FMA vector operations, OpenMP multi-threading, and gate fusion~\citep{OpenMP_1,OpenMP_2} to accelerate simulation speed and perform complex quantum circuit simulations. Integrated with Cirq~\citep{Cirq}, qsim is capable of executing simulations for up to 40 qubits on a 90-core Intel\textsuperscript{\textregistered} Xeon\textsuperscript{\textregistered} processor~\citep{IntelXeon}. Its performance is effectively showcased through its application in XEB~\citep{art19}.

These quantum software tools from Google Quantum AI~\citep{Cirq,TensorFlow_whitepaper,OpenFermion0,OpenFermion,qism} equip researchers with the necessary resources and capabilities to explore and expand the frontiers of quantum computing. By offering robust libraries such as Cirq~\citep{Cirq}, OpenFermion~\citep{OpenFermion0}, and  TensorFlow Quantum~\citep{TensorFlow_whitepaper}, Google enables researchers to explore the potential of quantum computing and push the boundaries of what is possible in solving complex problems.

\section{Grover search algorithm with Criq}

Grover's algorithm~\citep{Alg_Grover} is a quantum search algorithm designed to find a specific item from an unsorted database with quadratic speedup over classical search methods~\citep{GSA}. Given a black-box oracle function \( \xi (x) \) that marks the target solution \( x^\oplus  \) by flipping its phase, Grover's search algorithm ($\cal {GSA}$) identifies \( x^\oplus  \) with high probability. The algorithm operates in \( {\cal O}(\sqrt{N}) \) quantum operations, where \( N \) is the number of possible items, compared to the \( {\cal O}(N) \) operations required classically. The core of $\cal {GSA}$ involves the following steps:
\begin{enumerate}
     \item \textbf{Initialization:} Create a superposition of all possible states using Hadamard gates.
     \item \textbf{Oracle Query:} Apply the oracle to flip the phase of the state corresponding to the target solution.
     \item \textbf{Diffusion Operator:} Enhance the probability amplitude of the target state through inversion about the average amplitude.
     \item \textbf{Measurement:} Measure the quantum state to determine the solution.
\end{enumerate}
By iterating the oracle and diffusion operator, $\cal {GSA}$ converges to the solution with a high probability, making it an efficient tool for unstructured search problems. 
This section provides a demonstration of $\cal {GSA}$ using the Cirq framework. This demonstration adapted from~\citep{Cirq}. The implementation includes the initialization of qubits, construction of the oracle, application of the Grover operator, and measurement of the results.

$\bullet$ \textbf{Initialization of qubits:} 
Set up the qubits required for the algorithm. This includes initializing input qubits and an output qubit.
\begin{table}[H]
    \centering
    \begin{tabular}{c}
    \includegraphics[width=0.48\textwidth]{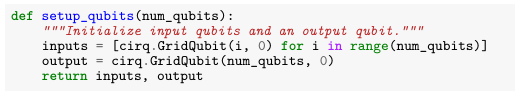} \\
    \end{tabular}
\end{table}

$\bullet$ \textbf{Oracle construction:} 
Create an oracle function that marks the correct solution by flipping the sign of the amplitude of the state that matches the target bit sequence.
\begin{table}[H]
    \centering
    \begin{tabular}{c}
    \includegraphics[width=0.48\textwidth]{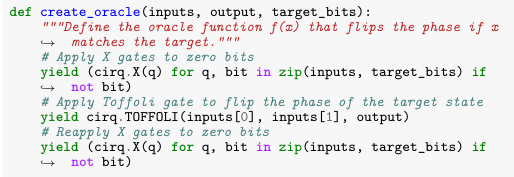} \\
    \end{tabular}
\end{table}

$\bullet$ \textbf{Grover operator implementation:} 
Implement the Grover operator, which includes initialization, applying the oracle, and performing amplitude amplification.
\begin{table}[H]
    \centering
    \begin{tabular}{c}
    \includegraphics[width=0.48\textwidth]{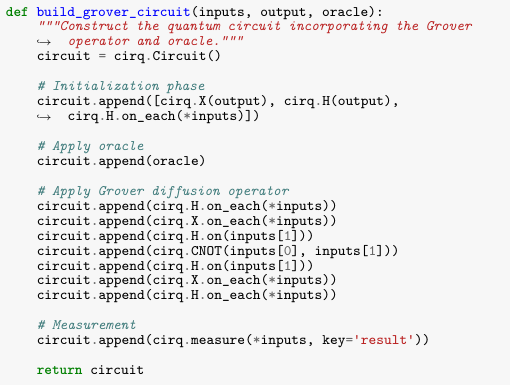} \\
    \end{tabular}
\end{table}

$\bullet$ \textbf{Simulation and measurement:} 
Simulate the quantum circuit and measure the results. Analyze the measurement results to find the most frequent outcome.
\begin{table}[H]
    \centering
    \begin{tabular}{c}
    \includegraphics[width=0.48\textwidth]{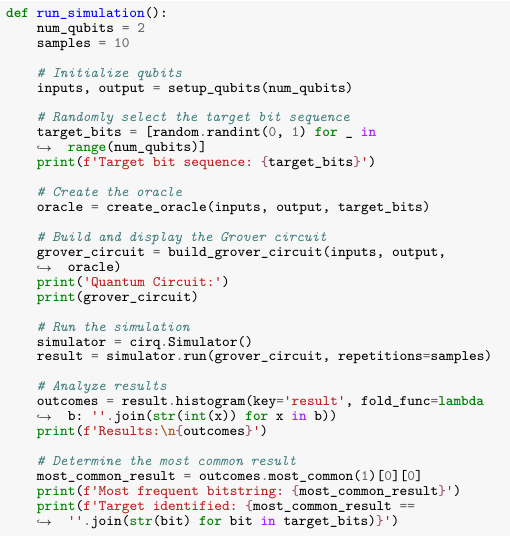} \\
    \end{tabular}
\end{table}

$\bullet$ \textbf{Result analysis:} 
Examine the results from the simulation to determine if the algorithm successfully identified the target bit sequence. 
\begin{table}[H]
    \centering
    \begin{tabular}{c}
    \includegraphics[width=0.48\textwidth]{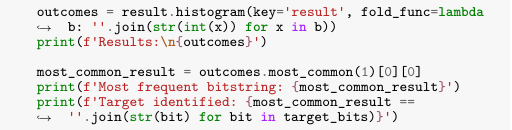} \\
    \end{tabular}
\end{table}

The output of the $\cal {GSA}$ is as follows: The target bit sequence identified by the algorithm was 
$[1,0]$. Upon simulation, the results showed a counter with a frequency of {`10': 10}, indicating that the bitstring `10' was observed in all measurement repetitions. The most frequent bitstring was therefore `10', which matched the target sequence. This confirms that the $\cal {GSA}$ successfully identified the target bitstring $[1,0]$ as the most common result from multiple circuit runs. A comprehensive characterization of $\cal {GSA}$ on state-of-the-art large-scale superconducting quantum hardware is presented in~\citep{GSA}. End-to-end experiments conducted on Google's quantum processors, utilizing algorithms and experiments that can be run with ReCirq~\citep{ReCirq}, a GitHub repository 
dedicated to research code that extends and leverages Cirq~\citep{Cirq}.

\section{Towards large-scale useful quantum computers}\label{sec:roadmap}

\begin{figure*}
    \centering
    \includegraphics[width=\textwidth]{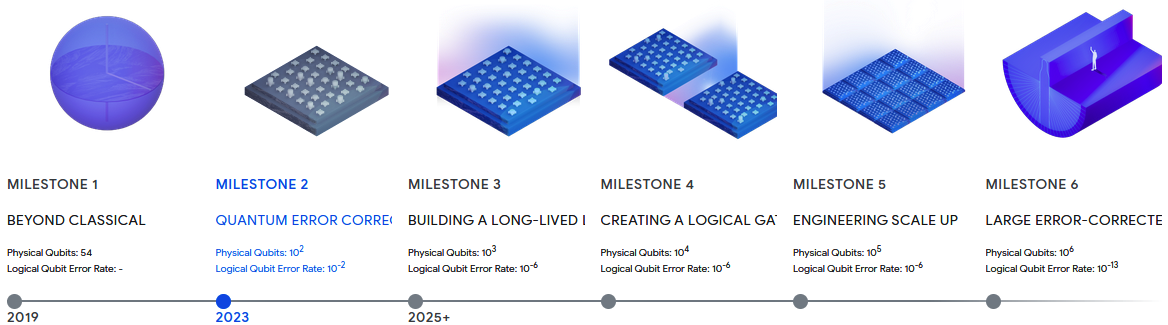}
\caption{The roadmap for Google Quantum AI's quantum computing. Aiming to fully realize the potential of this technology by creating a large-scale quantum computer that can handle complex computations with error correction. This journey is guided by six key milestones designed to advance their quantum computing hardware and software towards impactful applications. Image source~\citep{roadmap}.}
    \label{fig:roadmap2}
\end{figure*}

Google's Quantum AI aims to unlock the full potential of quantum computing by developing a large-scale quantum computer capable of complex, error-corrected computations. A roadmap for Quantum Computing at Google Quantum AI is shown in Figure~\ref{fig:roadmap2}. Their roadmap outlines six key milestones towards achieving high-quality quantum hardware and software for impactful applications~\citep{roadmap}:

\begin{enumerate}

    \item \textbf{Beyond classical (2019)} (physical qubits: 54): This milestone marked the achievement of quantum computational advantage. Google's Quantum AI team demonstrated this with their 53-qubit (SYC-53) quantum processor~\citep{art19}, which completed a specific computation (see Section~\ref{subsec:SYC53}) in 200 seconds—an endeavor that would have taken classical supercomputers 10,000 years~\citep{art19}. This result signifies a pivotal advancement in quantum computing capabilities beyond classical limitations.

    \item \textbf{Error-corrected qubits (2023)} (physical qubits: $10^2$, logical qubit error rate: $10^{-2}$): Achieving reliable quantum computing necessitates the use of error-corrected qubits, or logical qubits~\citep{QErrorCorrection1,QErrorCorrection2,GoogleAI2023_2}. In 2023, Google Quantum AI successfully demonstrated a prototype for a logical qubit, showing that errors could be mitigated by employing quantum error correction techniques~\citep{Google_superm_2023}. This achievement marks a transition from theoretical to practical quantum error correction, laying the groundwork for large-scale, functional quantum computers.

    \item \textbf{Building a long-lived logical qubit (2025+)} (physical qubits: $10^3$, logical qubit error rate: $10^{-6}$): This milestone focuses on developing logical qubits that can maintain coherence over long durations, performing up to one million operations with minimal errors. Achieving this involves scaling up error correction, enhancing qubit performance, and expanding infrastructure. The goal is to dramatically reduce error rates while increasing qubit count, with a requirement for meaningful computational benchmarks.

    \item \textbf{Creating a logical gate} (physical qubits: $10^4$, logical qubit error rate: $10^{-6}$): The development of logical gates is crucial for executing reliable quantum computations. This stage involves demonstrating low-error gates between logical qubits, akin to the universal gates in classical computers. Successfully creating these gates will advance the realization of practical error-corrected quantum applications.

    \item \textbf{Engineering scale-up} (physical qubits: $10^5$, logical qubit error rate: $10^{-6}$): This milestone aims at scaling quantum systems to incorporate more logical qubits with high-fidelity gate operations. The objective is to enable the implementation of several error-corrected quantum computing applications, marking a significant step toward practical and scalable quantum computing.

    \item \textbf{Large error-corrected quantum computer} (physical qubits: $10^6$, logical qubit error rate: $10^{-13}$): The ultimate milestone targets the development of a quantum computer with the capability to manage and control 1,000,000 qubits. This ambitious goal seeks to push the boundaries of quantum technology, with the potential to revolutionize various fields, including medicine and sustainable technology. Upon reaching this milestone, Google Quantum AI anticipates unveiling over 10 error-corrected quantum computing applications.
    
\end{enumerate}

\section{Conclusion}\label{Sec:Conclusion}

Quantum computing stands at the precipice of transforming technology, offering computational capabilities that promise to reshape industries and scientific endeavors. In this review, we have traced the remarkable journey of Google Quantum AI over the past decade, highlighting its pivotal role in advancing quantum computing technology. From its early developmental milestones to its groundbreaking achievements in quantum supremacy, Google Quantum AI’s achievements underscore the transformative potential of quantum technology and its capacity to address complex computational problems that are currently beyond the reach of classical systems. Through these pioneering efforts, Google Quantum AI has consistently pushed the boundaries of what is possible, in the realm of quantum computation, and paving the way for groundbreaking discoveries and applications.

Throughout the years (2013-2024), Google Quantum AI has demonstrated significant progress in various areas, including quantum hardware development, error correction, and software innovations. The evolution from initial experimental setups to the sophisticated quantum processors like \textit{Foxtail}, \textit{Bristlecone}, \textit{Weber}, \textit{Sycamore} (SYC-53, SYC-67 and SYC-70) and the ambitious quest for error-corrected quantum computers reflects a commitment to overcoming the fundamental challenges of quantum computing. Their work in extending qubit coherence times, reducing error rates, and enhancing computational capabilities has paved the way for more practical and scalable quantum systems.

Google Quantum AI's approach to evaluating effective quantum volume has provided valuable insights into the computational costs associated with RCS, OTOC experiments, and recent Floquet evolution studies. Although these experiments have not yet reached the stage of practical computational applications, Google Quantum AI is optimistic that advancements in error rates will enable OTOC experiments to achieve the first truly practical, beyond-classical applications of quantum processors.

As we look towards the future, Google Quantum AI’s continued efforts in improving quantum error correction, expanding quantum hardware, and refining quantum algorithms will be crucial in realizing the vision of large-scale, fault-tolerant quantum computers. While a full-scale error-corrected quantum computer has the potential to tackle problems beyond the reach of classical computers, developing such a device remains a formidable challenge and is still several years away.

\begin{acknowledgments}

The views and conclusions expressed in this paper are solely those of the author and do not necessarily represent the views of Google Quantum AI, NASA, the Universities Space Research Association, IBM Quantum, UC Berkeley, or any affiliated organization.

\end{acknowledgments}

\section*{Funding}

The author declares that no funding, grants, or other forms of support were received at any point throughout this research.

\end{document}